\documentclass[12pt]{article}
\usepackage{graphicx}
\usepackage{adjustbox}
\usepackage{setspace}
\usepackage{array}
\usepackage{booktabs}
\usepackage{colortbl}
\usepackage{enumerate}
\usepackage{longtable}
\usepackage{ragged2e}
\usepackage{siunitx}
\usepackage{threeparttable}

\newcolumntype{P}[1]{>{\raggedright\arraybackslash}p{#1}}

\usepackage{newtxtext,newtxmath} 
\usepackage{amsmath}
\usepackage{mathtools}
\usepackage{bm}
\usepackage[dvipsnames]{xcolor}
\usepackage{authblk}
\usepackage[a4paper, margin=1in]{geometry}

\usepackage{rotating}
\usepackage{multirow}

\usepackage[
    sorting=none,
    url=false,
    isbn=false,
    eprint=false,
    doi=false,
    bibstyle=numeric,
    citestyle=numeric-comp
]{biblatex}

\renewbibmacro{in:}{%
  \ifentrytype{article}{}{\printtext{\bibstring{in}\intitlepunct}}}

\AtEveryBibitem{%
  \clearfield{pages}%
  \clearlist{language}%
}

\usepackage[title]{appendix}

\title{A Bayesian Longitudinal Model for Imputing Item-Level Missing Data in Trial-Based Economic Evaluations}
\author[1,2]{Xiaoxiao Ling}
\author[3]{Andrea Gabrio}
\author[1]{Gianluca Baio}
\affil[1]{Department of Statistical Science, University College London, United Kingdom}
\affil[2]{Nuffield Department of Primary Care Health Sciences, University of Oxford, United Kingdom}
\affil[3]{Department of methodology and statistics, Faculty of Health Medicine and Life Science, Maastricht University, Maastricht, Limburg, Netherlands}
\date{}

\begin{document}

\maketitle

\section*{Abstract}
Trial-based economic evaluations are widely used to assess the cost-effectiveness of healthcare interventions and inform decision-making. Cost and effectiveness outcomes are typically collected using multi-item questionnaires administered at multiple time points, and are often subject to item-level missingness. In principle, imputation (i.e., replacing missing value with estimated or substituted values) should be performed at the item level to fully exploit available information. However, this is rarely implemented in practice due to several statistical challenges, including the longitudinal data structure, cross-item dependence, heterogeneous missingness patterns, and the mixture of skewed cost and count data. In this paper, we develop a Bayesian longitudinal model for imputing item-level missing data in trial-based economic evaluations that accommodates these complexities within a unified framework. The approach combines a transition-model formulation for longitudinal dependence with flexible distributional assumptions and explicit modelling of cross-item relationships, allowing item-level responses of different types to be coherently modelled over time. Motivated by a real-world trial, we demonstrate the flexibility and practical applicability of the proposed approach. We further discuss how the model can be extended to settings where data may be missing not at random.

\noindent \textbf{Keywords:} Bayesian methods; missing data; missing items; longitudinal data; cost-effectiveness analysis

\newpage

\section{Introduction}
Health economic evaluations alongside clinical trials, most commonly in the form of cost–effectiveness analysis (CEA), are widely used in countries with health technology assessment (HTA) agencies, particularly the UK, Canada, and Australia, to assess whether an intervention provides good value for money relative to relevant comparators. Typically, cost and utility data are collected at multiple time points using patient-reported resource use questionnaires \parencite{beecham_costing_2001, byford_cost-effectiveness_1999}, and generic quality-of-life instruments \parencite{the_euroqol_group_euroqol_1990, brazier_estimation_2002, brazier_estimation_2004}. These responses are aggregated over time by attaching unit costs and value sets, respectively, to calculate total costs and quality-adjusted life-years (QALYs) for each study arm. Then incremental costs and QALYs are derived to compute the incremental cost-effectiveness ratio (ICER), defined as the additional cost per QALY gained. Decision-making is informed by comparing the ICER and its associated uncertainty with prespecified willingness-to-pay (WTP) thresholds: interventions with ICERs below the thresholds are generally considered cost-effective.

Trial-based economic evaluations are frequently affected by missing outcome data, particularly in trials with lengthy resource use questionnaires and repeated measurements over multiple time points. Missingness often arises at the item level, where participants fail to complete individual questionnaire items, leading to missing total costs or QALYs. In principle, methods that make use of all available information should be used by addressing missingness at the item level. However, in practice, analysts often handle missing data at the level of most aggregated outcomes (i.e., total costs and QALYs) for ease of implementation \parencite{ling_scoping_2022}. This approach may discard useful information in observed items and overlook important data features in the analysis, such as the longitudinal structure of cost and utility measurements. Consequently, it may lead to inefficient or biased inference, particularly when missingness patterns vary across items and time.

From a statistical perspective, methods for handling item-level missing data are well established \parencite{simons_multiple_2015, rombach_multiple_2018, mainzer_comparison_2021, vera_is_2021, rosel_what_2022}. However, their application in cost-effectiveness analysis presents additional challenges due to the complexity of the data. Several features must be considered, including correlations between longitudinal item responses as well as dependence within items at the same time point. In addition, while aggregated cost-effectiveness outcomes are typically continuous, item-level data may include a mixture of data types, such as continuous and count variables. Recent work has highlighted the importance of modelling item-level resource use data within a Bayesian framework and comparing results across different levels of aggregation using empirical data \parencite{gabrio_bayesian_2026}. These approaches primarily focus on flexible joint modelling of outcomes and do not explicitly formulate cross-item dependence within its regression framework, nor do they provide guidance on resource use item selection or directly address extensions to informative missingness mechanisms.

To address these challenges, we adopt an explicit Bayesian modelling approach based on a transition model, which allows the conditional mean of each response to depend on both covariates—including other item responses—and previous observations, thereby capturing serial dependence while retaining a regression-based structure \parencite{zeger_markov_1988}. We explore the parametric flexibility of this approach using a motivating case study and extend the framework to account for informative missingness under a Missing Not At Random (MNAR) assumption. The core methodology and its application to the case study were previously presented in \textcite{ling_msr15_2022} and \textcite{ling_impact_2024}, and further developed in the author’s doctoral thesis \parencite{ling_item-level_2025}.

\section{Motivation: The ORBIT Trial}
This research was motivated by the ORBIT trial --- a multi-center randomised controlled trial conducted in two sites of England that compared the cost-effectiveness of the new intervention, online-delivered therapist-supported Exposure and Response Prevention (ERP) (n=112), with the control, online education (n=112), in children and young people with Tourette syndrome or chronic tics disorder from a health and social care cost perspective over 6 months \parencite{hollis_therapist-supported_2021}. Resource use data were collected using an adapted version of the Child and Adolescent Service Use Schedule (CA-SUS) at baseline, 3 and 6 months \parencite{byford_cost-effectiveness_1999, harrington_randomised_2000, barrett_mental_2006}, and were combined with unit costs from the Personal Social Services Research Unit (PSSRU) to estimate the costs of health and social care services at corresponding time points \parencite{curtis_unit_2019}. Total costs were calculated as the sum of healthcare and intervention costs for both arms. Utility data were collected using the CHU-9D instrument \parencite{furber_validity_2015}, a generic preference-based measure designed for children and adolescents aged 7–17 years, at baseline, 3, and 6 months. Responses at each time point were converted into utility values using an algorithm developed by \textcite{stevens_valuation_2012}, and were combined to calculate QALYs using an area-under-the-curve approach \parencite{manca_estimating_2005}.

In the original health economics analysis, cost data were ``cleaned" based on the missingness of the primary clinical outcome --- tic severity at 3 months, measured by the the Yale Global Tic Severity Scale (YGTSS-TTSS) \parencite{leckman_yale_1989}. This approach entailed that cost items were marked as ``missing” if the primary outcome was missing, and imputed as zero otherwise. Items left blank or zero were also considered ``missing” if the primary outcome was missing. This approach assumed the same missing data patterns for both clinical and cost outcomes, since both were collected concurrently with research assistance. When the primary outcome was present, missing cost data was assumed to reflect data entry errors rather than true non-response. However, this assumption does not align with the standard definition of missing data --- i.e., data intended to be collected but unavailable at the time of analysis. Therefore, cost data requires recleaning based on the proper definition of missing data rather than completion of primary outcome. For all analyses in these study, cost data were re-cleaned without relying on the original analysis' assumption. Details of the data recleaning process are provided in the Appendix~\ref{appendix_strategy}.

\subsection{Summary of Missingness in the ORBIT Trial}
The number, proportion and type of missingness and complete cases within each treatment arm are summarised in Appendix~\ref{appendix_proportion}. The amount of missing data increases steadily over time in both treatment arms for both cost-effectiveness outcomes. Surprisingly, health care costs suffer severe nonresponses from the baseline: approximately half of the patients are missing at baseline. The proportions of missingness increase to 67.9\% and 68.8\% at 6 months in the intervention and control arm, respectively. By the end of the follow-up, there are only 35 complete cases, defined as patients who have complete data for both costs and utilities at all three time points. The amounts of missingness are similar between the two arms, though the intervention arm tends to have slightly higher proportion of missing utility scores. 

The number of participants in the intervention arm, whose missing data are entirely due to missing items, almost remains stable across the three time points, while the number in the control arm fluctuates over time, from 59/112 (52.7\%) at baseline to 53/112 (47.3\%) at 3 months and 58/112 (51.8\%) at 6 months. 

\begin{figure}[p]
       {\centering
       \includegraphics[width=16cm]{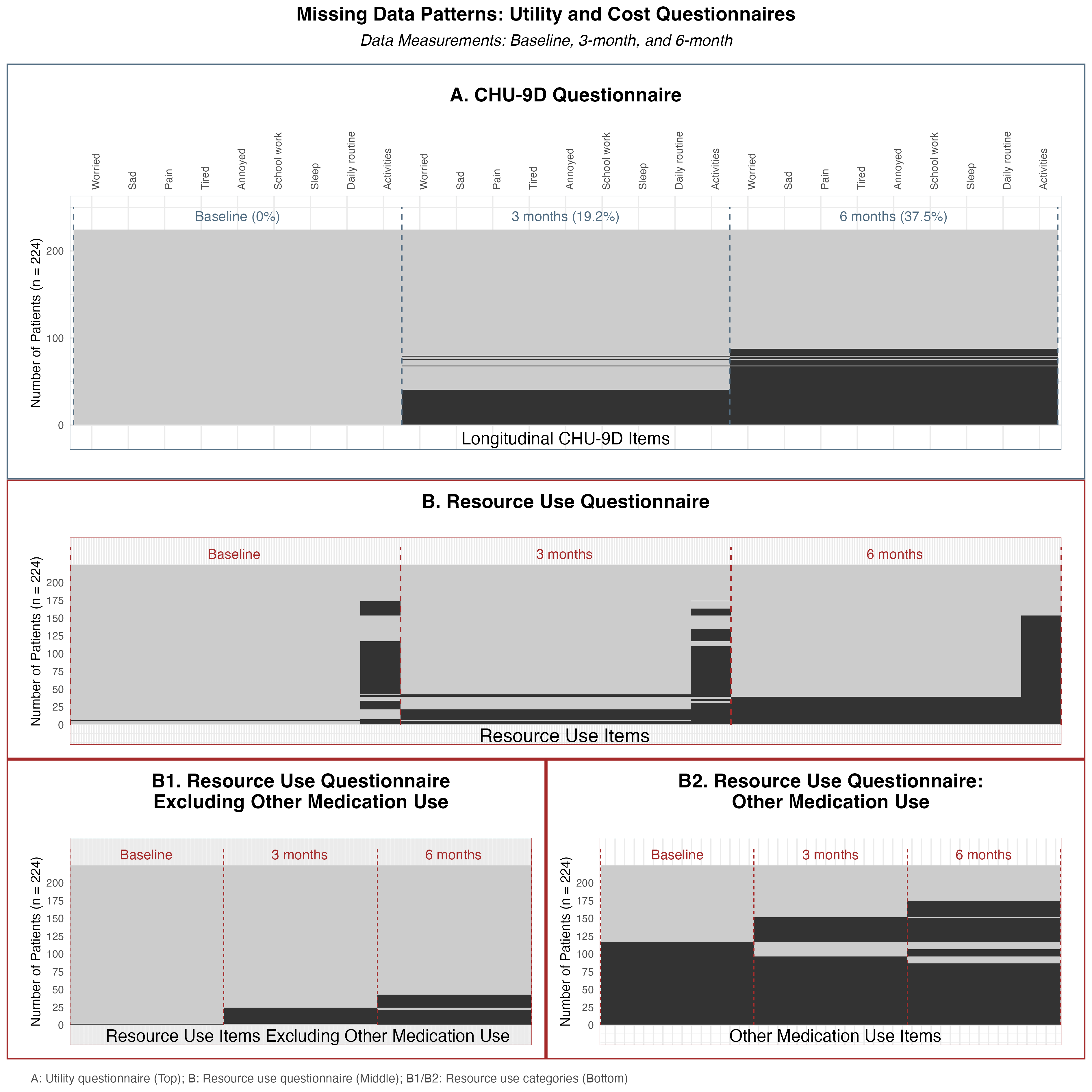}
       \caption{Missing patterns of the ORBIT trial data}
       \label{fig1}
       }
       {\small Panel A reports missing data patterns for the CHU-9D quality-of-life questionnaires over time, while Panel B presents missing data patterns for the resource use questionnaires. To illustrate differences in missingness across cost categories, selected cost categories are shown in Panel B1, while other medication use is presented in Panel B2. In each panel, grey areas represent observed values, while black areas indicate missing values.}
\end{figure}

Cost data exhibit a combination of both completely and partly missing questionnaires, the latter reflecting item-level missingness, whereas QALY data are exclusively affected by completely missing questionnaires (see Figure~\ref{fig1}). Item-level missingness mostly arises in resource use questions concerning other medication use for all three time points, making it the main cause for the missingness in total costs. 

\subsection{Item Selection}
Addressing item-level missingness in the cost questionnaires presents unique complexities. In the ORBIT trial, the adapted version of the CA-SUS questionnaire consists of 480 questions, posing a distinct challenge in developing a model for each individual cost item. Arguably, a more plausible approach for this particular scenario is to identify and focus on key cost categories upon which to develop the modelling approach.

A strategy is proposed to streamline the number of items in the cost questionnaire to a practical level by identifying a feasible subset of items that may be suitable for item-level modelling (see details in Appendix~\ref{appendix_items}). This strategy assumes that the ideal items for item-level modelling should exhibit a non-negligible level of missingness, be reported in a consistent scale over time, and present a relatively small proportion of zeros among observed values. Although developed specifically for this case study, the strategy provides a pragmatic approach to reducing the scope of the questionnaire.

Based on this strategy, three cost categories --- Child and Adolescent Mental Health Services at clinic (CAMHS), GP at clinic and Special Education Needs Co-ordinator (SENCO) --- have been identified as suitable for item-level modelling. These categories show a non-negligible amount of missing data, are reported consistently across three time points, and demonstrate a moderate frequency of zeros in the observed data: the proportion of zero values among observed items for the CAMHS and GP at clinic remains under 80\% over time. As for the SENCO, the proportion of zero items is 72.2\% and 76.6\% at baseline and 3 months respectively, though it rises to 81.1\% by the end of the follow-up period. The three identified cost categories are used to exemplify the item-level modelling that could be applied to more general scenarios. 

Not all cost categories are eligible to be modelled at the item level. For instance, other medication use exhibits a higher degree of missingness compared to the three identified cost categories and will be modelled as continuous cost data because the data are inconsistent, with people reporting their medication dosages using different scales and units of measure. Similar to the other medication use, the remainder of the cost questionnaire will be added up as remaining costs and modelled as continuous costs in an aggregated way.

\section{Conceptual Framework: Missing Cost Items and Utility Scores}
\subsection{Notation}
Let us consider a hypothetical trial-based CEA that compares the cost-effectiveness of $T$ treatments for $N$ participants at $J$ time points. For each treatment arm at each time point, a vector of response variables --- representing $K$ cost categories and a utility score --- is measured for every individual and denoted by $\bm{y}_{ijt}$, where $i = 1, \ldots, N; j = 0,\ldots,J; t = 1, \ldots, T$. Let $\bm{c}_{ijt} = (c_{ijt1},\ldots,c_{ijtK})$ denote the $K$ cost categories, each represented by $c_{ijtk}$, for $k = 1,\ldots,K$, and let $u_{ijt}$ denote the utility score. The vector of responses $\bm{y}_{ijt}$ can therefore be expressed as: 
\begin{equation}
  \bm{y}_{ijt} = (\bm{c}_{ijt}, u_{ijt}) = (c_{ijt1},\ldots,c_{ijtK}, u_{ijt})
\end{equation}
Total costs and QALYs over time are obtained using the costs $\bm{c}_{ijt}$ and utility scores $u_{ijt}$ measured at each time point $j$. 

\subsection{A Transition-Model Framework for Cost Item Modelling}
The primary focus of the item-level modelling framework is the joint distribution of multivariate responses, denoted by $\bm{y}_{it} = (\bm{y}_{i0t},\ldots,\bm{y}_{iJt})$, over the entire trial period:
\begin{equation}
    p(\bm{y}_{i0t}, \ldots, \bm{y}_{iJt}) = p(\bm{c}_{i0t}, u_{i0t}, \ldots, \bm{c}_{iJt}, u_{iJt})
\end{equation}

A \textit{transition model} is adopted to account for the longitudinal data structure \parencite{zeger_markov_1988}. This approach maintains a regression-based structure while simultaneously capturing serial dependence in longitudinal outcomes through a \textit{generalised linear model} framework. Specifically, the model defines the conditional mean of a response as a function of covariates and the history of past responses, allowing the dependence of current outcomes on both covariates and previous observations to be represented within a single equation \parencite{zeger_overview_1992}. A brief introduction to the transition model formulation is provided in the Appendix~\ref{appendix_transition}.

Given its strengths in capturing longitudinal dependence while preserving a mean regression structure, the transition model provides a natural foundation for item-level modelling in CEAs. In practice, a first-order Markov assumption is adopted, whereby the current response depends only on the most recent observation. The adaptation of the transition model in this context can be expressed as:
\begin{align}
     p(\bm{y}_{i0t}, \ldots, \bm{y}_{iJt}) & = p(\bm{y}_{i0t}) \prod_{j=1}^J p(\bm{y}_{ijt} \mid \bm{y}_{i(j-1)t}) \\
     h^{-1} \{\text{E}(\bm{y}_{ijt} | \bm{y}_{i(j-1)t})\} & = \alpha f(\bm{y}_{i(j-1)t}) + [\ldots]
\end{align}
where $h^{-1}\{.\}$ is the link function, $f(.)$ stands for the function of the history of past responses, $\alpha$ is the coefficient describing the effects of past responses on the current outcome, and the term $[\ldots]$ denotes the potential inclusion of covariates in the model.

Cost and utility data are typically correlated but may follow different distributional assumptions. To account for this correlation while ensuring flexibility in modelling each outcome, the following product formulation of the joint distribution is employed.

At each time point $j$, the joint conditional distribution of cost categories and utilities, $p(\bm{y}_{ijt} \mid \bm{y}_{i(j-1)t}) = p(\bm{c}_{ijt}, u_{ijt} \mid \bm{y}_{i(j-1)t})$, is factorised as the product of a conditional distribution of cost categories given utilities at the same time point as well as outcomes at the previous time point, $p(\bm{c}_{ijt} \mid u_{ijt}, \bm{y}_{i(j-1)t})$, and a conditional distribution of utilities given the outcomes at the previous time point, $p(u_{ijt} \mid \bm{y}_{i(j-1)t})$:
\begin{equation}
\begin{aligned}
    p(\bm{y}_{ijt} \mid \bm{y}_{i(j-1)t}) = & p(\bm{c}_{ijt}, u_{ijt} \mid \bm{y}_{i(j-1)t}) \\
    = & p(\bm{c}_{ijt} \mid u_{ijt}, \bm{y}_{i(j-1)t}) p(u_{ijt} \mid \bm{y}_{i(j-1)t}) 
\end{aligned}
\label{eq:product_formulation}
\end{equation}
This model structure accounts for the dependence between cost categories and utilities at the same time point and enables outcome-specific modelling, while also capturing potential serial dependence between the outcomes at different time points. Although alternative specifications of the factorisation are possible, in Equation~\ref{eq:product_formulation}, costs are conditioned on utilities due to the assumption that patients' health condition, measured by utilities, primarily drives the costs associated with their consumption of health care services. 

To allow a higher degree of flexibility when modelling each cost category, the conditional distribution, $p(\bm{c}_{ijt} \mid u_{ijt}, \bm{y}_{i(j-1)t})$, can be further factorised into a series of conditional distributions: 
\begin{equation}
\begin{aligned}
    p(\bm{c}_{ijt} & \mid u_{ijt}, \bm{y}_{i(j-1)t}) = \\
    & p(c_{ijt1} \mid u_{ijt}, \bm{y}_{i(j-1)t})
    \prod_{k=2}^K
    p(c_{ijtk} \mid c_{ijt(k-1)}, \ldots, c_{ijt1}, u_{ijt}, \bm{y}_{i(j-1)t})
\end{aligned}
\end{equation}
Non-Normal distributional assumptions can then be applied to different cost categories and utility scores to reflect the skewness and appropriate distributional assumptions of the data \parencite{gabrio_full_2019}. 

The proposed modelling framework can effectively address the complexities in item-level modelling for CEAs: the longitudinal data structure is reflected using the transition model; the potential correlation between costs and utilities is considered via the product formulation based on the underlying assumption that patients' health condition can influence the health care costs; and the flexibility of modelling cost category data is expressed through the use of different distributional assumptions.

\section{Resource Use Modelling Strategy}
\subsection{Practical Modelling Considerations}
Out of the 224 patients in the ORBIT trial, only 35 had fully observed data for other medication use over the entire follow-up period and were therefore complete cases, defined as patients with complete cost and utility data at all three time points. Meanwhile, a subset of cost categories have substantially more complete observations: 132 patients have complete data for the four cost categories (CAMHS, GP, SENCO, and remaining costs). 

To make full use of the available data while maintaining model interpretability, we develop three separate submodels: 
\begin{itemize}
    \item \textbf{Joint model of identified cost categories and utilities:} The four primary cost categories, including the three selected cost categories to be modelled at item level (i.e.~CAMHS, GP, SENCO) and the remaining aggregated costs, are modelled jointly with utilities. This joint model is factorised as the product of sequential conditional distributions of each cost category, alongside a distribution of utilities, thereby capturing the correlation between outcomes at the same time point.
    \item \textbf{Model for other medication use:} Other medication use is modelled separately, under the assumption of conditional independence from the other cost categories given utility values. This assumption reflects a compromise between modelling complexity and the small number of complete cases available for the other medication use.
    \item \textbf{Model for intervention costs:} Intervention costs, which were nearly complete except for two participants, are modelled independently of the other components.
\end{itemize}
Covariates for these models were initially selected following the original health economics analysis, with additional missingness predictors being identified (see Appendix~\ref{appendix_predictors}). We included site for the four primary cost categories, age and comorbidity for utilities, as well as site and comorbidity for other medication use.

\begin{figure}[p]
       {\centering
       \includegraphics[width=14cm]{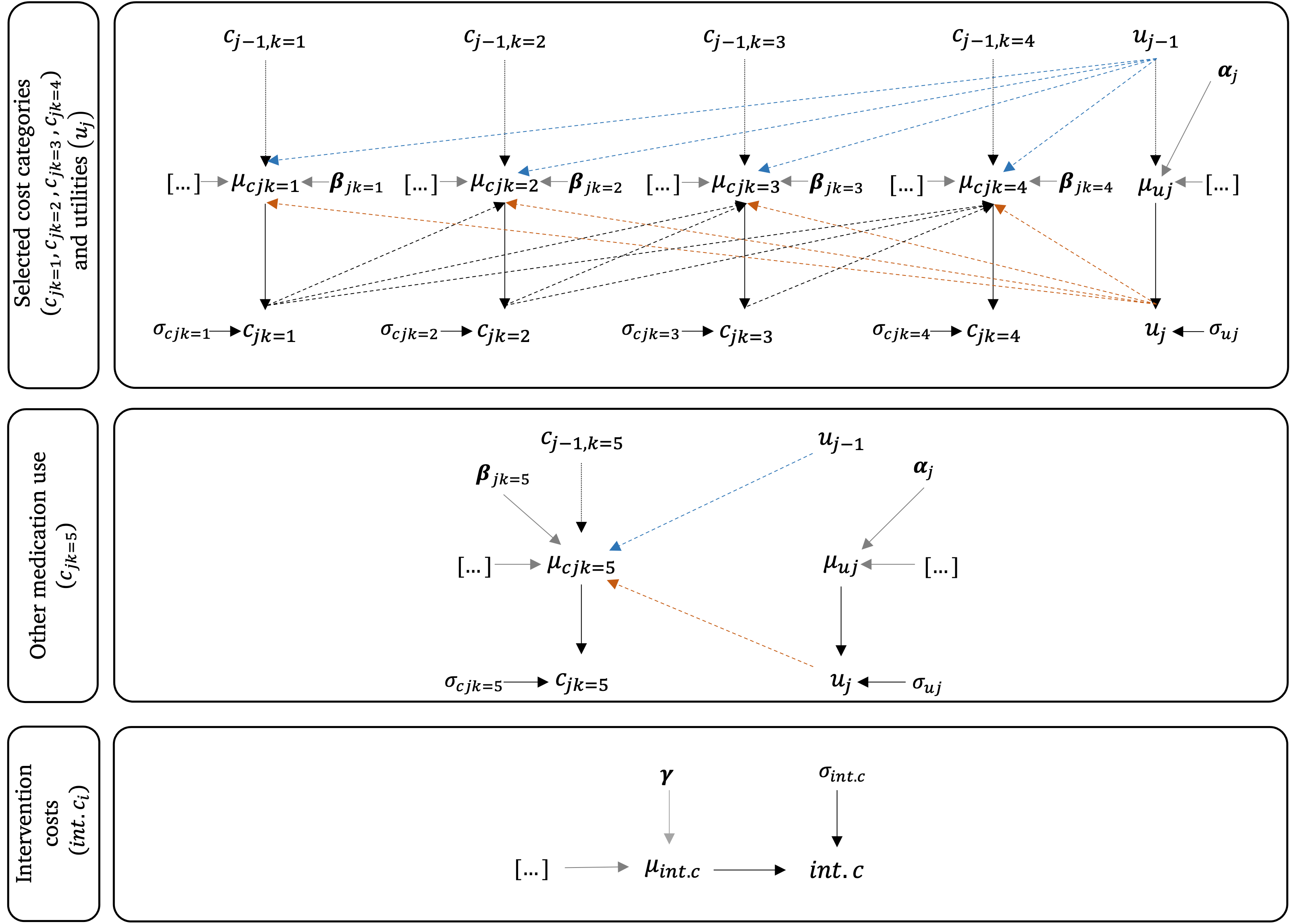}
       \caption{Graphic representation of item-level modelling for the ORBIT trial}
       \label{fig2}
       }
       {\small The framework consists of three model components, representing the model for the four primary cost categories and utilities, other medication use, and intervention costs, respectively. The letter $j$ denotes the time point $j \geq 1$ while the letter $k$ denotes the $k$th cost category. The Greek letters represent the parameters indexing each distribution. The black solid arrows show the relationship between these parameters and outcomes; the black dashed arrows suggest the dependence among resource use at the same time point; the black dotted arrows indicate the serial dependence between outcomes at different time points; the blue and orange dashed arrows represent resource use's dependence on utility values at the previous and current time points, respectively. The notation $[\ldots]$ denotes any potential inclusion of covariates whose relationships with mean parameters of outcome variables are indicated using gray solid arrows.}
\end{figure}

Figure~\ref{fig2} is a graphical representation of the modelling framework for the ORBIT trial, using three blocks representing (1) the model for the four primary cost categories and utilities; (2) the other medication use; and (3) the intervention costs, respectively. In this figure, individual and treatment indices $i$ and $t$ are omitted. The distribution of each outcome at each time point $j$ is indexed by a mean parameter (resource use: $\mu_{cjk}$; intervention costs: $\mu_{int.c}$; utilities: $\mu_{uj}$) and a standard deviation parameter (resource use: $\sigma_{cjk}$; intervention costs: $\sigma_{int.c}$; utilities: $\sigma_{uj}$). The relationships between the covariates and outcomes are quantified by vectors of coefficients: $\bm{\beta}_{jk}$ in the resource use model, $\bm{\alpha}_{j}$ in the utility model and $\bm{\gamma}$ in the intervention cost model.

\subsection{Joint Modelling of Cost Categories and Utilities}
Let $\bm{c}_{ijt} = (c_{ijt1}, \ldots, c_{ijt4})$ denote the four primary cost categories 
(CAMHS, GP, SENCO, and remaining costs) for $k = 1, \ldots, 4$, and let $u_{ijt}$ denote utility at time point $j \geq 1$. The individual and treatment indices $i$ and $t$ are omitted to ease notation. 

The joint distribution at time point $j$ is factorised as the product of two components. The first model specifies the distribution of cost categories for $\bm{c}_{j}$, conditional on utility values at the same time $u_{j}$, and on costs $\bm{c}_{(j-1)}$ and utility $u_{(j-1)}$ at the previous time point $j-1$: 
\begin{equation*}
    p(\bm{c}_{j} |u_{j}, \bm{c}_{j-1}, u_{j-1})
\end{equation*}

The second model specifies the distribution of utility values ($u_j$) at time point $j$, conditional on cost categories ($\bm{c}_{j-1}$) and utility values ($u_{j-1}$) at the previous time point $j-1$:
\begin{equation*}
    p(u_{j} | \bm{c}_{j-1}, u_{j-1}) 
\end{equation*}

The first cost model component is further decomposed into a series of conditional distributions of each cost category $k$: 
\begin{equation}
\begin{aligned}
    p(\bm{c}_{j} & \mid u_{j}, \bm{c}_{j-1}, u_{j-1}) = \\ 
    & p(c_{j1} \mid u_{j}, \bm{c}_{j-1}, u_{j-1}) \prod_{k=2}^4 p(c_{jk} \mid c_{j(k-1)}, \ldots, c_{j1}, u_{j}, \bm{c}_{j-1}, u_{j-1}) 
\end{aligned}
\end{equation}

The small sample size of the trial makes model implementation challenging and can lead to convergence issues. To achieve a balance between model complexity and convergence, the implemented model adopts a simplified specification of the general framework in the following respects: 
\begin{enumerate}[(1)]
    \item The cost category at current time point ($c_{jk}$) only depends on the same cost category ($c_{(j-1)k}$) and utility values at the previous time point ($u_{j-1}$), as well as other cost categories ($\bm{c}_{j,-k}$) and utility values ($u_{j}$) at the same time point. This adjustment, i.e., maintaining serial dependence within each cost category and in utilities, is motivated by the assumption that a patient’s current health state is typically strongly correlated with their previous state. In contrast, individual cost components are more likely to be influenced by the patient’s health status or previous use of specific services rather than by other health care expenses incurred at the previous time point. Therefore, it is assumed that the current cost category ($c_{jk}$) does not depend on other cost categories at the previous time point ($\bm{c}_{(j-1),-k}$); 
    \item Utility values at the current time point ($u_{j}$) depend only on utility values ($u_{j-1}$) at the previous time point. Removing cost categories from the utility model aligns with the underlying assumption that patients’ health condition drives their consumption of health care resources, rather than the opposite --- a feasible but less likely scenario.
\end{enumerate}

\subsection{Model for Other Medication Use}
The other medication use is modelled independently from other cost categories. The joint distribution of other medication use ($c_{jk=5}$) and utility values ($u_{j}$) at time point $j \geq 1$ is parameterised as a conditional distribution of other medication use given utilities at the same time point, other medication use ($c_{(j-1)k=5}$) and utilities ($u_{j-1}$) at the previous time point:
\begin{align*}
    p(c_{jk=5} \mid u_{j}, c_{(j-1)k=5}, u_{j-1})
\end{align*}

and a distribution of utilities at time point $j$, $p(u_{j})$. Ideally, the distribution of utilities at the current time point should be conditional on utilities at previous time point $j-1$, consistent with its counterpart in the first model component of the four primary cost categories, $p(u_{j} \mid u_{j-1})$. However, utilities are modelled conditional only on completely observed covariates, due to the following reasons: (1) the limited sample size of 35 cases for this model component constrains the adaptation of the more complex utility model from the first component, potentially leading to convergence issues; (2) utilities, serving as covariates in the other medication use model and presenting missing data at later time points, are required to be imputed but are not the primary focus of this model component. 

\subsection{Model for Intervention Costs}
The intervention costs are modelled similarly to the other medication use but do not need to be considered for multiple time points, since they are costed as a cross-sectional measure. In other words, there is only one model for the intervention costs across the entire follow-up.

\section{Full Bayesian Specification}
Different distributional assumptions have been explored for modelling the costs (i.e.,~the four primary cost categories, other medication use, and intervention costs) and utility values at each time point. These distributions were chosen based on literature and specific data features observed \parencite{mihaylova_review_2011, basu_regression_2012}. The model starts with assuming both resource use costs and utility values are normally distributed, as a basis for comparison. Next, to address the data skewness, the resource use costs are modelled using a Gamma distribution and utility values are modelled via a Beta distribution. 

Alternative distributional choices were considered by modelling three of the four primary cost categories (CAMHS, GP, and SENCO) as resource use count, where the resulting marginal mean resource use count was combined with the corresponding unit cost of the service to compute resource use costs. A Poisson model with fixed effects was initially developed by modelling CAMHS, GP, and SENCO service use as count data while retaining a Gamma distribution for the remaining costs. This model was later expanded to a Poisson model with random effects to overcome the limitation of the former, i.e.~equidispersion, defined as the mean of the random variable being equal to its variance. 

The linear predictors used across models are consistent, differing only in their distributional assumptions, link functions, and transformations of covariates. Therefore, the regression structure is defined primarily under the Normal model, and only model-specific modifications are described for the other models. The distributional assumptions and link functions for the item-level models are summarised in Table~\ref{table1}.

\subsection{Continuous Cost Model}
In this section, we illustrate the modelling specifications assuming cost categories as continuous cost data. To ease presentation of the models, we exemplify the implementation of our framework using the SENCO costs $c_{ijk=1}$ and utility values $u_{ij}$ at time point $j \geq 1$ only, due to the fact that the model specification would be similar across cost categories and model components. 

\subsubsection{Bivariate Normal Model} \label{method:normal}
\textit{Cost Model:} The SENCO ($c_{ijk=1}$) costs, for individual $i$ at time point $j$, can be modelled as: 
\begin{align*}
    & c_{ijk=1} \mid u_{ij} \sim \text{Normal}(\mu_{cijk=1}, \sigma^2_{cjk=1}) \\
    & \begin{aligned}
        \mu_{cijk=1} & = \beta_{0,j,k=1} + \beta_{1,j,k=1} \times \text{site}_i + \beta_{2,j,k=1} \times u^{*}_{ij} \\ 
                     & + \beta_{3,j,k=1} \times u^{*}_{i(j-1)} + \beta_{4,j,k=1} \times c^{*}_{i(j-1)k=1} 
    \end{aligned}
\end{align*}
where $\mu_{cijk=1}$ denotes the individual-level mean SENCO costs at time point $j$, and $\sigma_{cjk=1}$ represents their conditional standard deviation. The expressions, $u^{*}_{ij} = (u_{ij} - \bar{u}_{j})$, $u^{*}_{i(j-1)} = (u_{i(j-1)} - \bar{u}_{(j-1)})$, and $c^{*}_{i(j-1)k=1} = (c_{i(j-1)k=1} - \bar{c}_{(j-1)k=1})$, denote the mean-centred utilities at the current time point, utilities and SENCO costs at the previous time point, respectively. A vector of regression parameters, $\bm{\beta}_{j,k=1} = (\beta_{0,j,k=1}, \beta_{1,j,k=1}, \beta_{2,j,k=1}, \beta_{3,j,k=1}, \beta_{4,j,k=1})$, represents the intercept, effects of site, utilities at the current time point, utilities and SENCO costs at the previous time point, respectively.

\textit{Utility Model:} Utility values at time point $j$ follow a Normal distribution with individual-level mean $\mu_{uij}$ and conditional standard deviations $\sigma_{uj}$. Recall that the utility values at time point $j$ only depend on the utilities at the previous time point $j-1$, with age and comorbidity as other covariates, independent of any cost categories:
\begin{align*}
    u_{ij} \mid u_{i(j-1)}  & \sim \text{Normal}(\mu_{uij}, \sigma^2_{uj}) \\
    \mu_{uij} & = \alpha_{0,j} + \alpha_{1,j} \times \text{age}^{*}_i + \alpha_{2,j} \times u^{*}_{i(j-1)} + \alpha_{3,j} \times \text{cm}_i 
\end{align*} 
The age of each participant $i$ has been centred around its mean as a covariate in the model and is denoted by $\text{age}^{*}_i = (\text{age}_i - \bar{\text{age}})$. The term $\text{cm}_i$ denotes the comorbidity. $\bm{\alpha}_j = (\alpha_{0,j}, \alpha_{1,j}, \alpha_{2,j}, \alpha_{3,j})$ represents the regression parameters.

\textit{Prior Specification:} The models at different time points are completed by allocating suitable prior distributions to the parameters: $\text{Normal}(0,100^2)$ is given to the regression coefficients in all models at all times, and $\text{Uniform}(0,1000)$ is assigned to standard deviations of costs and utilities at different time points. 

\subsubsection{Beta-Gamma Model} \label{method:gamma}
To reflect the skewness of cost and utility data, a Gamma distribution is considered for costs and a Beta distribution is applied for utilities. Since the Gamma distribution requires data to be strictly positive, a small constant $\eta = 0.1$ has been added to all the original cost data before rescaling. The chosen value $\eta = 0.1$ is chosen to be small relative to the scale of the original data, minimising its impact on the overall distribution. 

\textit{Cost Model:} The cost model structure follows that defined in Section~\ref{method:normal}, with differences arising from the choice of link functions and transformations of covariates. The SENCO costs at time point $j \geq 1$, denoted by $c_{ijk=1}$, are now modelled as: 
\begin{align*}
        & c_{ijk=1} | u_{ij}, c_{i(j-1)k=1}, u_{i(j-1)} \sim \text{Gamma}(\mu_{cijk=1} \tau_{cijk=1}, \tau_{cijk=1}) 
\end{align*}
where $\mu_{cijk=1}$ is the individual-level mean of SENCO costs at time point $j$ and $\tau_{cijk=1}$ is the corresponding individual-level rate parameter. 

Any cost at the previous time point used as a covariate in the current cost model is transformed into logarithmic form due to the use of the log link function in the Gamma model. However, cost categories often have missing data, which may be imputed as small values close to zero in their imputation models. The logarithm of these imputed values can potentially reach negative infinity, leading to convergence issues in the model of other cost categories. 

To solve this issue, considering the small constant $\eta = 0.1$ has been added to the original cost data and the data have then been re-scaled by dividing by 10, the imputed values are theoretically never smaller than 0.01. Therefore, any cost covariate with missing data in the Beta-Gamma model, $c_{ijk}$, is transformed as $\Tilde{c}_{ijk} = \max(\log(0.01),\log(c_{ijk}))$. This transformation ensures that the logarithm of imputed values, which could approach infinity, is replaced with a reasonable value, and avoids impacting those imputed values that are close to zero but have finite logarithmic values.

\textit{Utility Model:} Utility values are modelled using a Beta distribution at time point $j$ where $j \geq 1$. The Beta distribution is defined by two shape parameters, $\kappa_{uij} = \mu_{uij}\phi_{uij}$ and $\gamma_{uij} = (1-\mu_{uij})\phi_{uij}$, where $\mu_{uij}$ represents the individual-level mean, while $\phi_{uij}$ is the individual-level scale parameter, defined as $\phi_{uij} = \frac{(1-\mu_{uij})\mu_{uij}}{\sigma_{uj}^2} - 1$ based on the conditional standard deviation of utilities at time point $j$ ($\sigma_{uj}$). Such parameterisation allows us to place priors on more intuitive parameters in the CEA context, specifically the standard deviations of the utilities. 

The utility values at time point $j$ are modelled together with but independently of the four primary cost categories. The model can be expressed as: 
\begin{align*}
       & u_{ij} \mid u_{i(j-1)} \sim \text{Beta}(\mu_{uij}\phi_{uij}, (1-\mu_{uij})\phi_{uij}) 
\end{align*}

\textit{Prior specification:} Normal prior distributions have been placed on the regression coefficients: $\text{Normal}(0,100^2)$ for the cost model and $\text{Normal}(0,3^2)$ for the utility model (due to the use of logit link). For standard deviations on both costs and utilities at different time points, Uniform prior distributions have been used: $\text{Uniform}(0,1000)$ for cost standard deviation and $\text{Uniform}(0,\sqrt{\mu_{uj}(1-\mu_{uj})})$ for utility standard deviation. 

\subsection{Resource Use Count Model}
The resource use of the three selected cost categories (SENCO, GP, and CAMHS) at each time point $j$ for individual $i$, denoted by $n_{ijk}$, where $k = 1,2,3$, is modelled as count data using a Poisson distribution. The mean resource use count is linked to its linear predictors through a log link function and then combined with their corresponding unit cost to calculate category-specific costs. The models for the utility values remain consistent with their respective components in the Beta-Gamma model and are omitted here to avoid duplication.

\subsubsection{Poisson Model with Fixed Effects} \label{method:poisson_fixed}
In the Poisson model with fixed effects, the resource use count is transformed into its logarithm as a covariate in the count data models rather than the logarithm of the resource use costs as used in Section~\ref{method:gamma}. To avoid the negative infinity arising with the use of logarithm transformation when there is a zero value, the term $\hat{n}_{ijk} = \log(\max(\theta, n_{ijk}))$ is used where $\theta = 0.5$ \parencite{zeger_markov_1988, li_random-effects_2007, lee_marginalized_2019}. The choice of $\theta = 0.5$ is supported by a series of sensitivity analyses over a plausible set of values (i.e.~0.1, 0.2, ..., 0.5), where $\theta = 0.5$ resulted in the lowest DIC, indicating the best model fit. Higher values beyond 0.5, though theoretically feasible, have been excluded because they are less distinct from observed values of 1 in the original data, and tend to inflate incremental mean costs (see Appendix~\ref{appendix_poisson}).

\textit{Cost Count Model:} Taking the number of SENCO contacts ($n_{ijk=1}$) at time point $j$ as an example, it is modelled conditional on utilities ($u_{ij}$) at the same time point, the number of SENCO contacts ($n_{i(j-1)k=1}$) and utility values ($u_{i(j-1)}$) at previous time point $j-1$. The number of contacts with SENCO ($n_{ijk=1}$) is modelled as:  
\begin{align*}
    & n_{ijk=1} | u_{ij}, n_{i(j-1)k=1}, u_{i(j-1)} \sim \text{Poisson}(\mu_{nijk=1})\end{align*}
where $\mu_{nijk=1}$ is the individual-level mean and variance of SENCO contacts at time point $j$. 

\textit{Prior specification:} Priors remain the same as those in Section~\ref{method:gamma}. $\text{Normal}(0,100^2)$ has been assigned to regression coefficients in the cost model while $\text{Normal}(0,3^2)$ has been given to those in the utility model. Uniform prior distributions have been considered for standard deviations: $\text{Uniform}(0,1000)$ on cost standard deviation and $\text{Uniform}(0,\sqrt{\mu_{uj}(1-\mu_{uj})})$ on utility standard deviation.

\subsubsection{Poisson Model with Random Effects} \label{method:poisson_random}
To relax the strict equidispersion constraint of the Poisson model with fixed effects, a possible solution is to introduce a random intercept, denoted by $\lambda_{ijk}$, to the model for each selected resource use $k$ at time point $j$. By allowing this intercept to vary at the individual level $i$, the model can better account for the inherent variability of the resource use data. Apart from using the random intercept, the Poisson model with random effects remains identical to Section~\ref{method:poisson_fixed}.

\textit{Prior specification:} Priors in Section~\ref{method:poisson_fixed} have been maintained. Additionally, the random intercepts $\lambda_{ijk}$ are assumed to have a Normal distribution, $\text{Normal}(0, \sigma^2_{\lambda jk})$. The standard deviation $\sigma_{\lambda jk}$ is given a minimally-informative Uniform prior, $\text{Uniform}(0,1)$.

\section{Application}
\subsection{Implementation}
To effectively use the available data, the models for the three selected cost categories and remaining costs have been constructed based on 132 samples with fully observed data for these cost categories and utility values at all three time points. There are only 35 participants with fully observed other medication use data, thus the model for the other medication use could only be developed using the 35 cases.

Results of item-level imputation on the original cost scale were initially reported under the missing at random (MAR) assumption, broadly assuming that missingness depended only on observed data \parencite{rubin_inference_1976}. We explored the robustness of the results in sensitivity analyses.

All the models were fitted using \texttt{JAGS} \parencite{plummer_jags_2003}, a program for Bayesian inference based on Markov Chain Monte Carlo (MCMC), via the \texttt{R2jags} package \parencite{su_r2jags_2021}, which the software to be called from within an R session (using R version 4.5.2). For all the models, two chains were run, each with 95,000 iterations and a burn-in of 15,000 iterations. A thinning rate at 16 was applied to decrease autocorrelation, leading to 10,000 iterations saved for inference. Convergence was evaluated for key model parameters using potential scale reduction statistic, $\hat{R}$ \parencite{gelman_inference_1992}: if the value was under 1.05, the model can be assumed not to have a convergence problem. The presence of possible issues in model convergence was also checked through a visual inspection of trace plots. 

Model fit was assessed by posterior predictive checks based on the data utilised for the model development \parencite{gelman_posterior_1996}. This dataset offers a more realistic representation of the ORBIT trial data compared with the complete cases which discards a significant portion of available information collected in observed items. The posterior predictive check was performed by comparing the distribution of replicated samples, drawn from posterior predictive distributions, with the distribution of observed data. The replicated data from a model with good fit can successfully capture the shape of observed data \parencite{gabry_visualization_2019}. 

The appropriateness of prior distributions was evaluated via prior predictive checks to ensure that these priors were minimally informative. In addition, alternative prior distributions were considered as a sensitivity analysis to check whether they can convey minimal impact on the results as intended. For instance, different values were considered for the upper bounds of the Uniform prior distributions on standard deviations. Posterior results of all models were robust to these prior specifications (Appendix~\ref{appendix_prior}).

\subsection{Model Assessment} 
Posterior predictive checks were used to assess how well the item-level models captured the distributional features of the observed data. Overall, clear differences emerged across model specifications. 

For continuous cost, the Normal model performed poorly, failing to capture key features such as the spikes and skewness observed in the empirical distributions. In contrast, the Beta-Gamma model provided a substantially better fit, with replicated datasets more closely reflecting the distributions of the observed data. 

When modelling resource use as count data, Poisson models generally provided an adequate fit across cost categories. However, limitations could be observed in some cases, particularly for SENCO counts at baseline, where the model with fixed-effects failed to reproduce the spike for small values. Allowing for random effects in this context improved model performance, yielding replicated distributions that more closely matched the observed data. We provide an illustration of posterior predictive checks for the four models in Appendix~\ref{appendix_ppc}.

\subsection{Missing Not at Random} 
We explored departures from MAR assumption for cost outcomes only, as the primary objective of this study was to investigate item-level modelling of resource use. 

We assessed the robustness of the results using a pattern-mixture modelling approach \parencite{little_pattern-mixture_1993}, in which a benchmark model under MAR was initially defined and then departures from the benchmark were introduced via sensitivity parameters. Specifically, we introduced a constant mean-score departure parameter ($\Delta_{s,k}$) across time points, representing a shift in linear predictors for individuals with missing-data pattern $s$ within cost category~$k$. The model can be specified as:
\begin{equation}
    g(\mu_{i,k,s}) = \bm{x}_{i,k} \beta_{\cdot,k} + \Delta_{s,k}
\end{equation}
where $\bm{x}_{i,k}$ denotes the covariate vector and $\beta_{\cdot,k}$ represent the corresponding regression coefficients. For the fully observed reference pattern ($S=0$), we set $\Delta_{s=0,k} = 0$.

The parameter $\Delta_{s,k}$ was assumed to be constant over time within each cost category to ease model implementation under the assumption of time-constant MNAR effects. The departure parameters were assigned Uniform priors to allow variation over a range of plausible values. We restricted the $\Delta_{s,k}$ to non-negative values because costs are non-negative and several cost categories included structural or near-structural zeros. For the Normal model (identity link), we specified $\text{Uniform}(0,2\sigma_{ck})$, where $\sigma_{ck}$ denotes the empirical standard deviation for cost category~$k$. The upper bound was chosen to calibrate departures relative to the observed variability \parencite{gabrio_bayesian_2020}. For models employing a log link (i.e.~Beta-Gamma and Poisson models), we applied $\text{Uniform}(0,2\log(\sigma_{ck}))$. 

Given that healthcare costs were partitioned into two broad components --- (1) the three selected cost categories (CAMHS, GP, SENCO), together with remaining costs, and (2) other medication use --- and each presents its unique missing-data patterns (Appendix~\ref{appendix_mnar}), MNAR sensitivity analyses were performed separately for each cost component. We report results for other medication use component in the two continuous cost models (i.e., the Normal and Beta–Gamma models), as the submodel for other medication use was specified identically in the Beta–Gamma and Poisson models. 

\subsection{Results} 
Table~\ref{table3} presents the cost-effectiveness results, reported as posterior means and 95\% credible intervals (CIs), under MAR. Mean QALYs were very similar across models in both treatment arms. All models produced positive incremental QALYs, ranging from 0.0041 (Poisson model with fixed effects) to 0.0058 (Normal model), implying modest health gains associated with the intervention over the 6-month follow-up period. However, 95\% CI for incremental QALYs included zero in all models, indicating substantial uncertainty around the estimated incremental QALYs and no clear evidence of a difference in QALYs between treatment arms.

Incremental cost estimates varied across distributional assumptions. The Beta–Gamma model yielded the lowest incremental cost (£65), whereas the Normal and Poisson models produced slightly higher but comparable estimates (Normal: £90; Poisson fixed effects: £74; Poisson random effects: £87). The 95\% CIs for incremental costs also included zero in all cases, reflecting considerable uncertainty regarding cost differences between treatment arms.

\begin{figure}[p]
       {\centering
       \includegraphics[width=14cm]{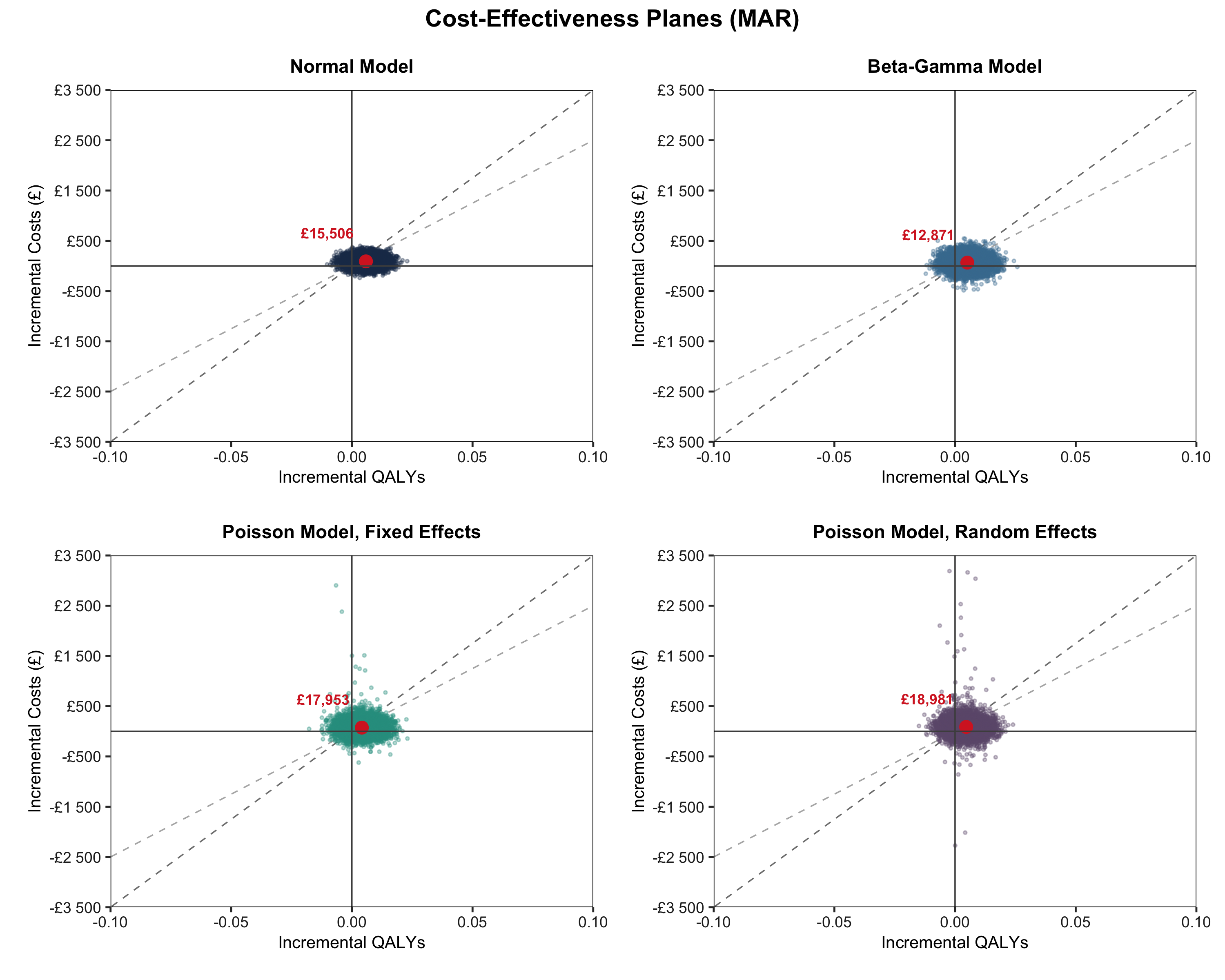}
       \caption{Cost-effectiveness planes of item-level models under MAR. MAR = Missing at random; QALY = quality-adjusted life-year. Each panel displays simulated incremental costs and QALYs. The dashed lines represent willingness-to-pay thresholds of £25,000 and £35,000 per QALY gained. The red point denotes the ICER estimated from each model.}
       \label{fig3}
       }
\end{figure}

The cost-effectiveness of the intervention was explored using pairs of population-level mean incremental costs and QALYs drawn from the posterior distributions at each MCMC iteration and graphically presented on cost-effectiveness planes (Figure~\ref{fig3}) \parencite{black_ce_1990}. The WTP thresholds of £25,000 and £35,000 per QALY gained --- values used by the UK governmental body, the National Institute for Health and Care Excellence (NICE), for new medicines from April 2026 --- were plotted as light and dark grey lines, respectively.

The Beta–Gamma and Poisson models exhibited greater uncertainty in incremental costs compared with the Normal model. In particular, the two Poisson models showed a wider vertical spread of posterior draws and included some extreme values in the upper half of the plane. This aligns with the findings in Table~\ref{table3}, where the uncertainty in incremental costs is higher in the two Poisson models compared with the two continuous cost models.

Across all models, most posterior draws fell in the north-east quadrant, indicating higher costs and improved health outcomes associated with the intervention. The majority of posterior draws fell below the corresponding threshold lines, suggesting that the intervention is very likely to be cost-effective compared with the control in all cases.

\begin{figure}[p]
       {\centering
       \includegraphics[width=14cm]{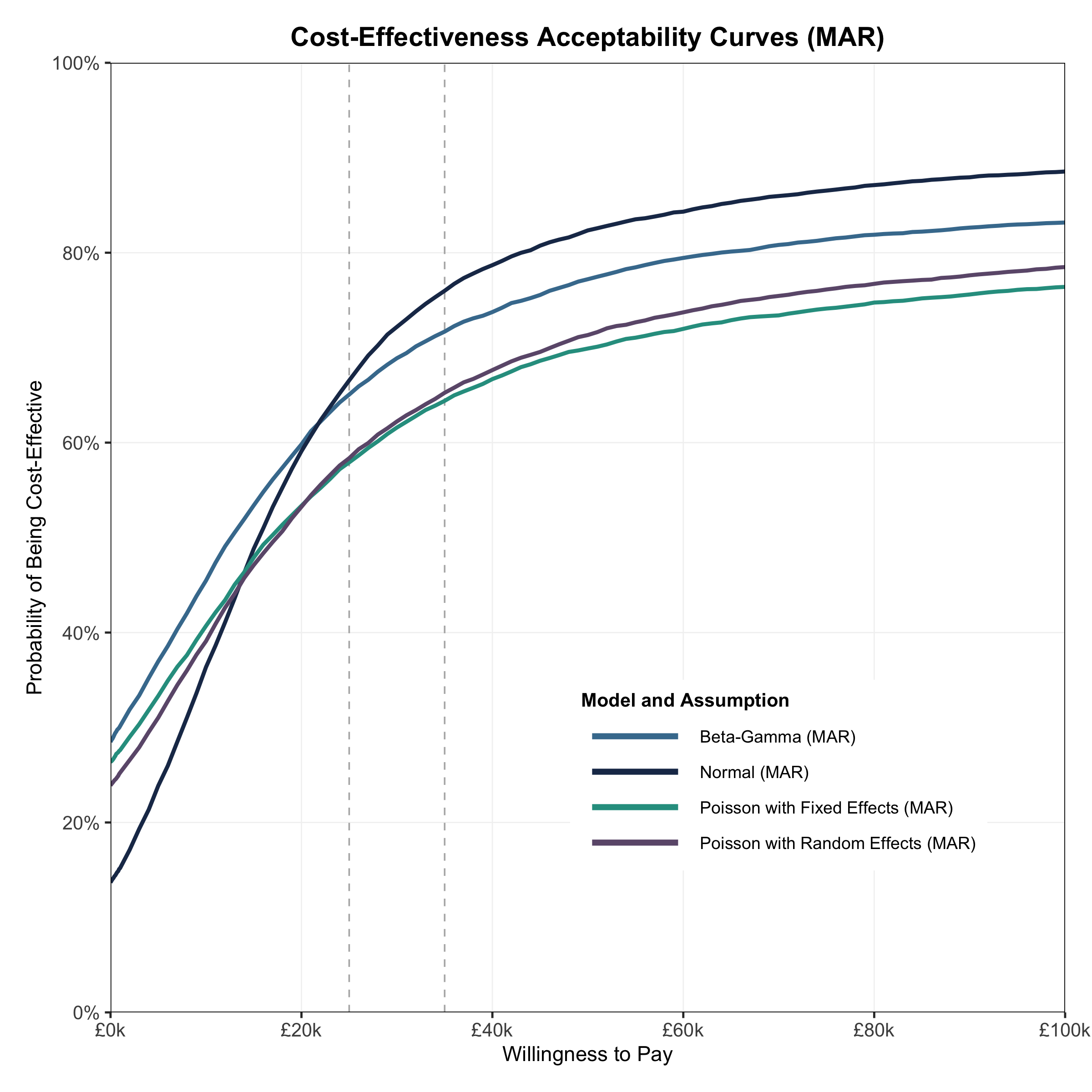}
       \caption{Cost-effectiveness acceptability curves for item-level models under MAR. MAR = Missing at random. The two vertical dashed lines indicate willingness-to-pay thresholds of £25,000 and £35,000 per QALY gained.}
       \label{fig4}
       }
\end{figure}

The probability that the intervention is cost-effective at different WTP thresholds is presented using cost-effectiveness acceptability curves (CEACs) \parencite{fenwick_guide_2005}, which illustrate the uncertainty in cost-effectiveness conclusions across different WTP values (Figure~\ref{fig4}). The solid lines represent the curves for the models under MAR. The two grey dashed vertical lines indicate WTP thresholds of £25,000 and £35,000 per QALY gained, respectively.

Overall, whenever modelling cost categories as costs or as resource-use counts, the probability of cost-effectiveness increases monotonically as the decision-maker’s WTP threshold increases. However, modelling choices influence the probability of the cost-effectiveness of the intervention. The intervention is less likely to be cost-effective under the Poisson models than under the two continuous cost models when the threshold exceeds £15,000. Between the two continuous cost models, the Normal model consistently leads to a higher probability of cost-effectiveness with respect to the Beta–Gamma model when the threshold goes beyond £20,000.

\subsection{Sensitivity Analysis}
\begin{figure}[p]
       {\centering
       \includegraphics[width=1\textwidth]{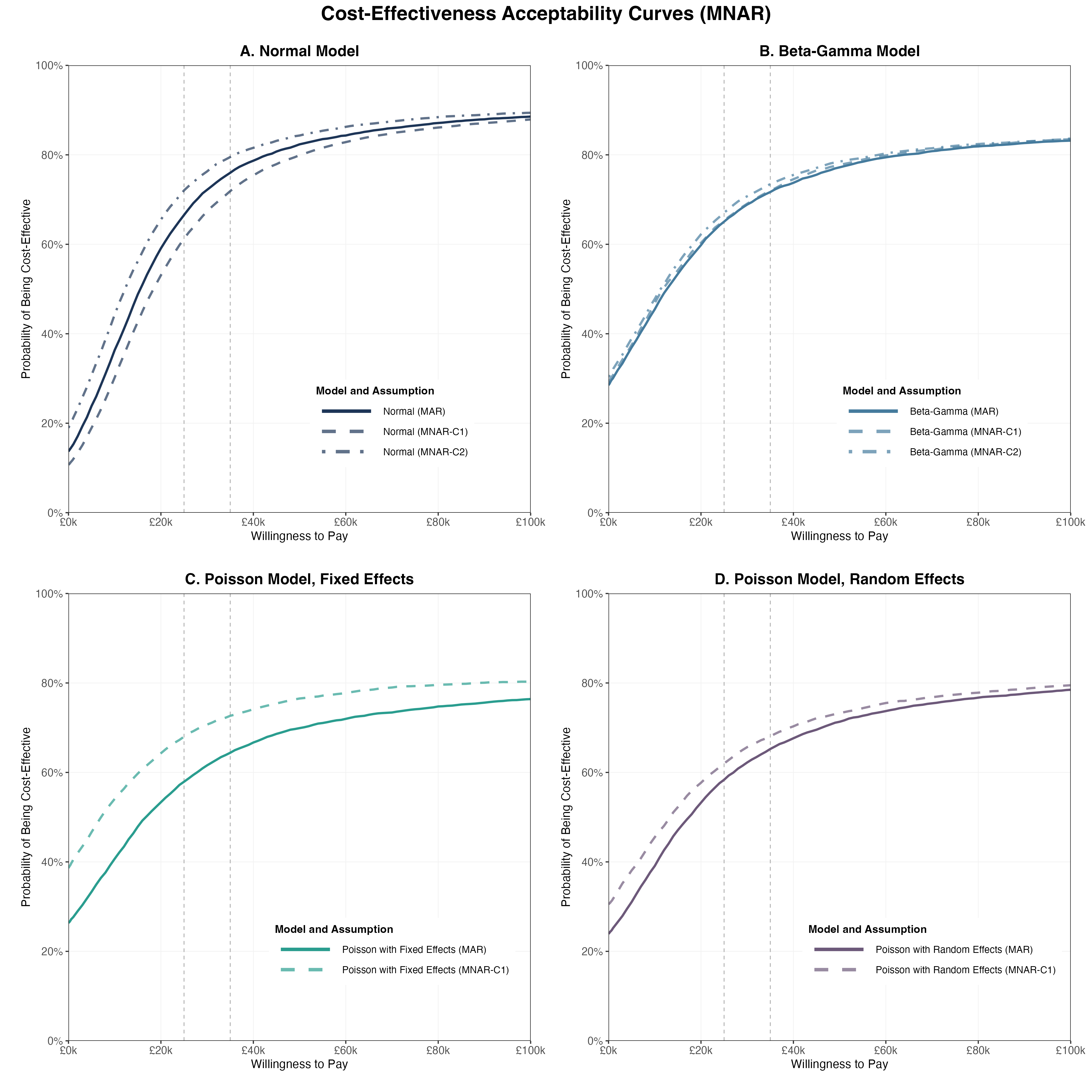}
       \caption{Cost-effectiveness acceptability curves for item-level models under MNAR. MAR = Missing at random; MNAR = Missing not at random. Solid curves represent models under MAR; dashed curves show sensitivity analyses under MNAR for the four primary cost categories (C1); and dot-dashed curves represent MNAR analyses for the other medication use (C2), conducted for the Normal and Beta–Gamma models only. The two vertical dashed lines indicate willingness-to-pay thresholds of £25,000 and £35,000 per QALY gained.}
       \label{fig5}
       }
\end{figure}

The CEACs under MNAR are shown in Figure~\ref{fig5}. Dashed lines represent the selected cost-category scenarios (C1), while dot–dashed lines correspond to the alternative specification for other medication use (C2). The cost-effectiveness planes for the non-ignorable scenarios are presented in Appendix~\ref{appendix_mnar} to save space.

The MNAR analyses show similar trends to those observed under MAR. However, the MNAR scenario for other medication use (C2; dot-dashed) consistently yielded higher probabilities of cost-effectiveness compared to the MAR analysis, highlighting the influence of missing-data assumptions on decision uncertainty. In particular, the Poisson model with fixed effects appears more sensitive to departures from MAR compared with the other three models, as reflected by the greater distance between their CEACs across assumptions.

\section{Discussion}
In this study, we introduced a Bayesian modelling framework based on a transition model for item-level imputation in trial-based CEAs, fully addressing its complexities including the correlation between multivariate outcomes, longitudinal data structure and the choice between modelling cost items as costs or resource use. We showed the feasibility of performing trial-based CEAs in the presence of missing items at the most disaggregated (item) level, and highlighted the methodological and practical challenges associated with such approach. Our empirical analysis overall suggested that the intervention is very likely to be cost-effective across most scenarios. However, the Normal and Beta-Gamma models presented higher probability of intervention being cost-effective than the Poisson models, suggesting how distributional assumptions may meaningfully influence economic conclusions.

The case study, the ORBIT trial, is representative of routine economic evaluations that are often affected by severe item-level missingness in cost questionnaires. Despite the specific data-processing procedures used and the missingness patterns observed, the trial provides a valuable test bed with potential for wider generalisability to routine economic evaluations, where sample sizes are often limited and missing-data definitions are not always rigorously applied during data cleaning.

In the original trial analysis, however, missing cost items were imputed as zero and some observed items were discarded when the primary clinical outcome was missing, followed by complete case analysis. Such way of handling missing data deviates from standard definitions of missing data in the statistical literature and does not make full use of the available information. While it is common for studies with questionnaires to handle missing items using simple imputation before the main analysis \parencite{fox-wasylyshyn_handling_2005, berchtold_treatment_2019}, this approach might cause biased results, especially when subject to item-level missingness and repeated for multiple cases \parencite{parent_handling_2013}.

Item-level modelling for cost items in CEAs may be carried out based on two options: modelling the selected cost items as costs for each service or as the resource use. In this study, the feasibility of both approaches has been explored. Cost items may have more spikes around certain values such as zeros than cost data at more aggregated levels, which may require special attention in model fitness assessment when using continuous distributions. On the other hand, modelling the cost items as resource use with Poisson models ensures consistency with most observed data. However, the choice of two modelling approaches does not result in much difference in cost-effectiveness conclusions in our case study, particularly under MAR. The models exhibit similar results in both cost-effectiveness planes and CEACs, possibly leading to same cost-effectiveness conclusions. This aligns with findings from one previously-published trial-based CEA that compares the impact of performing the imputation for each cost category as costs or resource use \parencite{sharples_amaze_2018}. 

Trial-based economic evaluations primarily focus on the estimation of marginal (population-level) mean cost and QALYs over time. For this reason, we specifically employed a transition model framework, which directly parameterises the conditional distribution of outcomes given previous observations while preserving the marginal mean structure of interest. Although alternative longitudinal models such as growth mixture models could in principle be considered \parencite{muthen_general_2002}, the three time points restricted the reliable identification of trajectory structure. 

While recent Bayesian modelling framework has been developed to address partially observed health resource use data for CEAs \parencite{gabrio_bayesian_2026}, our work has extended this literature in several directions. First, we formally integrated transition modelling with the analytical structure of trial-based health economic evaluations. Second, we comprehensively examined the impact of different distributional assumptions on inference and highlighted how modelling choices propagate into uncertainty around cost-effectiveness conclusions. Third, we exemplified how this proposed item-level model could be extended to accommodate MNAR settings using a pattern mixture approach specification and investigate its influence on real-world decision-making.

The proposed method has several limitations to acknowledge: first, the model was primarily designed to handle missing items in cost data alone. This focus was driven by practical considerations, as cost questionnaires are typically more lengthy than utility questionnaires in real applications, resulting in greater possibility to encounter CEAs with item-level missingness in costs only. Second, when identifying cost items suitable for item-level modelling, the unit cost of each service could be taken into account. Compared to items with small unit costs, indicating that the service use will exert minimal impact on total costs, it might be more efficient to focus on items with higher unit costs. However, this decision should be made together with health economists, trialists and clinicians to prevent overlooking cost items that are clinically or policy-relevant. In our case study, unit cost was not used to identify items due to the lack of relevant information. Lastly, sensitivity analyses for missing-data assumptions were restricted to relatively simple MNAR scenarios. This pragmatic choice was mostly driven by the limited sample size in the case study. Nevertheless, the framework is sufficiently flexible to accommodate more complex MNAR specifications in larger datasets. Future research could extend this work by developing and evaluating more comprehensive approaches such as selection models to modelling non-ignorable mechanisms for item-level missingness \parencite{mason_flexible_2021}.

In summary, we developed a Bayesian transition modelling framework to handle missing items within trial-based economic evaluations. By integrating item-level imputation with practical modelling challenges within a unified Bayesian structure, this work establishes a principled foundation for item-level handling in CEAs and provides a flexible platform for future research to explore extensions to more complex longitudinal designs and missing-data mechanisms.

\section*{Funding}
This work was supported by the Engineering and Physical Sciences Research Council Doctoral Training Partnership, UK (EP/R513143/1). The funding body had no role in the design of the study, interpretation of the data, preparation of the manuscript, or the decision to submit for publication.

\section*{Acknowledgements}
The authors would like to thank Prof. Rachael Hunter and Miss Marie Le Novere, health economists for the ORBIT trial at UCL, UK, for facilitating access to the trial data. The authors would also like to thank Prof. Gian Luca Di Tanna and Dr Menelaos Pavlou for their valuable comments on the thesis upon which this manuscript is based.

\printbibliography

@article{little_pattern-mixture_1993,
	title = {Pattern-mixture models for multivariate incomplete data},
	volume = {88},
	number = {421},
	journal = {Journal of the American Statistical Association},
	publisher = {Taylor \& Francis},
	author = {Little, Roderick JA},
	year = {1993},
	pages = {125--134},
}

@article{brazier_estimation_2002,
	title = {The estimation of a preference-based measure of health from the {SF}-36},
	volume = {21},
	number = {2},
	journal = {Journal of health economics},
	author = {Brazier, John and Roberts, Jennifer and Deverill, Mark},
	year = {2002},
	pages = {271--292},
}

@article{brazier_estimation_2004,
	title = {The estimation of a preference-based measure of health from the {SF}-12},
	journal = {Medical care},
	author = {Brazier, John and Roberts, Jennifer},
	year = {2004},
	pages = {851--859},
}

@article{zeger_overview_1992,
	title = {An overview of methods for the analysis of longitudinal data},
	volume = {11},
	copyright = {http://onlinelibrary.wiley.com/termsAndConditions\#vor},
	issn = {0277-6715, 1097-0258},
	url = {https://onlinelibrary.wiley.com/doi/10.1002/sim.4780111406},
	doi = {10.1002/sim.4780111406},
	language = {en},
	number = {14-15},
	urldate = {2025-03-10},
	journal = {Statistics in Medicine},
	author = {Zeger, Scott and Liang, Kung‐Yee},
	month = jan,
	year = {1992},
	pages = {1825--1839},
}

@article{leckman_yale_1989,
	title = {The {Yale} {Global} {Tic} {Severity} {Scale}: initial testing of a clinician-rated scale of tic severity},
	volume = {28},
	number = {4},
	journal = {Journal of the American Academy of Child \& Adolescent Psychiatry},
	publisher = {Elsevier},
	author = {Leckman, James F and Riddle, Mark A and Hardin, Maureen T and Ort, Sharon I and Swartz, Karen L and Stevenson, JOHN and Cohen, Donald J},
	year = {1989},
	pages = {566--573},
}

@article{gabrio_bayesian_2026,
	title = {A {Bayesian} {Modeling} {Framework} for {Health} {Care} {Resource} {Use} and {Costs} in {Trial}-{Based} {Economic} {Evaluations}},
	volume = {46},
	url = {https://doi.org/10.1177/0272989X251376026},
	doi = {10.1177/0272989X251376026},
	number = {2},
	journal = {Medical Decision Making},
	author = {Gabrio, Andrea},
	year = {2026},
	pages = {158--173},
}

@article{muthen_general_2002,
	title = {General growth mixture modeling for randomized preventive interventions},
	volume = {3},
	number = {4},
	journal = {Biostatistics},
	publisher = {Oxford University Press},
	author = {Muthén, Bengt and Brown, C Hendricks and Masyn, Katherine and Jo, Booil and Khoo, Siek-Toon and Yang, Chih-Chien and Wang, Chen-Pin and Kellam, Sheppard G and Carlin, John B and Liao, Jason},
	year = {2002},
	pages = {459--475},
}

@article{beecham_costing_2001,
	title = {Costing psychiatric interventions},
	volume = {2},
	journal = {Measuring mental health needs},
	author = {Beecham, Jennifer and Knapp, Martin and {others}},
	year = {2001},
	note = {Publisher: Gaskell: London},
	pages = {200--224},
}

@phdthesis{ling_item-level_2025,
	type = {{PhD} {Thesis}},
	title = {Item-{Level} {Imputation} in {Trial}-{Based} {Economic} {Evaluations}},
	school = {UCL (University College London)},
	author = {Ling, Xiaoxiao},
	year = {2025},
}

@article{ling_msr15_2022,
	title = {{MSR15} {A} {Bayesian} {Approach} for {Handling} {Missing} {Items} in {Trial}-{Based} {Cost}-{Effectiveness} {Analysis} with {Multi}-{Item} {Questionnaires}},
	volume = {25},
	number = {7},
	journal = {Value in Health},
	publisher = {Elsevier},
	author = {Ling, Xiaoxiao and Gabrio, Andrea and Mason, Alexina and Baio, Gianluca},
	year = {2022},
	pages = {S520},
}

@article{ling_scoping_2022,
	title = {A {Scoping} {Review} of {Item}-{Level} {Missing} {Data} in {Within}-{Trial} {Cost}-{Effectiveness} {Analysis}},
	issn = {10983015},
	url = {https://linkinghub.elsevier.com/retrieve/pii/S1098301522001115},
	doi = {10.1016/j.jval.2022.02.009},
	urldate = {2022-04-27},
	journal = {Value in Health},
	author = {Ling, Xiaoxiao and Gabrio, Andrea and Mason, Alexina and Baio, Gianluca},
	year = {2022},
}

@inproceedings{ling_impact_2024,
	address = {Berlin, Germany},
	title = {Impact of performing imputation at different level of missingness in trial-based cost- effectiveness analysis with multi-item questionnaires},
	volume = {44},
	url = {https://doi.org/10.1177/0272989X231225267},
	doi = {10.1177/0272989X231225267},
	publisher = {SAGE PUBLICATIONS INC},
	author = {Ling, Xiaoxiao and Gabrio, Andrea and Mason, Alexina and Baio, Gianluca},
	year = {2024},
	pages = {PP--123},
}

@article{stevens_valuation_2012,
	title = {Valuation of the {Child} {Health} {Utility} {9D} {Index}},
	volume = {30},
	issn = {1179-2027},
	url = {https://doi.org/10.2165/11599120-000000000-00000},
	doi = {10.2165/11599120-000000000-00000},
	number = {8},
	journal = {PharmacoEconomics},
	author = {Stevens, Katherine},
	month = aug,
	year = {2012},
	pages = {729--747},
}

@article{the_euroqol_group_euroqol_1990,
	title = {{EuroQol} - a new facility for the measurement of health-related quality of life},
	volume = {16},
	issn = {0168-8510},
	url = {https://www.sciencedirect.com/science/article/pii/0168851090904219},
	doi = {https://doi.org/10.1016/0168-8510(90)90421-9},
	number = {3},
	journal = {Health Policy},
	author = {The EuroQol Group},
	year = {1990},
	pages = {199--208},
}

@article{carpenter_missing_2021,
	title = {Missing data: {A} statistical framework for practice},
	volume = {63},
	issn = {0323-3847, 1521-4036},
	shorttitle = {Missing data},
	url = {https://onlinelibrary.wiley.com/doi/10.1002/bimj.202000196},
	doi = {10.1002/bimj.202000196},
	language = {en},
	number = {5},
	urldate = {2024-01-23},
	journal = {Biometrical Journal},
	author = {Carpenter, James and Smuk, Melanie},
	month = jun,
	year = {2021},
	pages = {915--947},
}

@article{rombach_multiple_2018,
	title = {Multiple imputation for patient reported outcome measures in randomised controlled trials: advantages and disadvantages of imputing at the item, subscale or composite score level},
	volume = {18},
	issn = {1471-2288},
	shorttitle = {Multiple imputation for patient reported outcome measures in randomised controlled trials},
	url = {https://bmcmedresmethodol.biomedcentral.com/articles/10.1186/s12874-018-0542-6},
	doi = {10.1186/s12874-018-0542-6},
	language = {en},
	number = {1},
	urldate = {2021-07-06},
	journal = {BMC Medical Research Methodology},
	author = {Rombach, Ines and Gray, Alastair and Jenkinson, Crispin and Murray, David and Rivero-Arias, Oliver},
	month = dec,
	year = {2018},
	pages = {87},
}

@article{zeger_markov_1988,
	title = {Markov {Regression} {Models} for {Time} {Series}: {A} {Quasi}-{Likelihood} {Approach}},
	volume = {44},
	issn = {0006341X},
	shorttitle = {Markov {Regression} {Models} for {Time} {Series}},
	url = {https://www.jstor.org/stable/2531732?origin=crossref},
	doi = {10.2307/2531732},
	language = {en},
	number = {4},
	urldate = {2023-05-27},
	journal = {Biometrics},
	author = {Zeger, Scott and Qaqish, Bahjat},
	month = dec,
	year = {1988},
	pages = {1019},
}

@article{heagerty_marginalized_2000,
	title = {Marginalized multilevel models and likelihood inference (with comments and a rejoinder by the authors)},
	volume = {15},
	number = {1},
	journal = {Statistical Science},
	author = {Heagerty, Patrick and Zeger, Scott},
	year = {2000},
	pages = {1--26},
}

@article{simons_multiple_2015,
	title = {Multiple imputation to deal with missing {EQ}-{5D}-{3L} data: {Should} we impute individual domains or the actual index?},
	volume = {24},
	issn = {0962-9343},
	url = {http://link.springer.com/10.1007/s11136-014-0837-y},
	doi = {10.1007/s11136-014-0837-y},
	number = {4},
	journal = {Qual. Life Res.},
	author = {Simons, Claire and Rivero-Arias, Oliver and Yu, Ly-Mee and Simon, Judit},
	month = apr,
	year = {2015},
	pages = {805--815},
}

@article{gelman_inference_1992,
	title = {Inference from {Iterative} {Simulation} {Using} {Multiple} {Sequences}},
	volume = {7},
	language = {en},
	number = {4},
	journal = {Statistical Science},
	author = {Gelman, Andrew and Rubin, Donald},
	year = {1992},
	pages = {457--472},
}

@article{rubin_inference_1976,
	title = {Inference and missing data},
	volume = {63},
	number = {3},
	journal = {Biometrika},
	author = {Rubin, Donald},
	year = {1976},
	pages = {581--592},
}

@article{mihaylova_review_2011,
	title = {Review of statistical methods for analysing healthcare resources and costs},
	volume = {20},
	issn = {1057-9230, 1099-1050},
	url = {https://onlinelibrary.wiley.com/doi/10.1002/hec.1653},
	doi = {10.1002/hec.1653},
	language = {en},
	number = {8},
	urldate = {2022-01-13},
	journal = {Health Economics},
	author = {Mihaylova, Borislava and Briggs, Andrew and O'Hagan, Anthony and Thompson, Simon},
	month = aug,
	year = {2011},
	pages = {897--916},
}

@article{manca_estimating_2005,
	title = {Estimating mean {QALYs} in trial-based cost-effectiveness analysis: the importance of controlling for baseline utility},
	volume = {14},
	issn = {1057-9230, 1099-1050},
	shorttitle = {Estimating mean {QALYs} in trial-based cost-effectiveness analysis},
	url = {https://onlinelibrary.wiley.com/doi/10.1002/hec.944},
	doi = {10.1002/hec.944},
	language = {en},
	number = {5},
	urldate = {2022-09-02},
	journal = {Health Economics},
	author = {Manca, Andrea and Hawkins, Neil and Sculpher, Mark},
	month = may,
	year = {2005},
	pages = {487--496},
}

@article{mason_flexible_2021,
	title = {Flexible {Bayesian} longitudinal models for cost‐effectiveness analyses with informative missing data},
	volume = {30},
	issn = {1057-9230, 1099-1050},
	url = {https://onlinelibrary.wiley.com/doi/10.1002/hec.4408},
	doi = {10.1002/hec.4408},
	language = {en},
	number = {12},
	urldate = {2022-02-12},
	journal = {Health Economics},
	author = {Mason, Alexina and Gomes, Manuel and Carpenter, James and Grieve, Richard},
	month = dec,
	year = {2021},
	pages = {3138--3158},
}

@article{faria_guide_2014,
	title = {A {Guide} to {Handling} {Missing} {Data} in {Cost}-{Effectiveness} {Analysis} {Conducted} {Within} {Randomised} {Controlled} {Trials}},
	volume = {32},
	issn = {1170-7690, 1179-2027},
	url = {http://link.springer.com/10.1007/s40273-014-0193-3},
	doi = {10.1007/s40273-014-0193-3},
	language = {en},
	number = {12},
	urldate = {2021-07-24},
	journal = {PharmacoEconomics},
	author = {Faria, Rita and Gomes, Manuel and Epstein, David and White, Ian},
	month = dec,
	year = {2014},
	pages = {1157--1170},
}

@article{gabrio_bayesian_2020,
	title = {A {Bayesian} parametric approach to handle missing longitudinal outcome data in trial‐based health economic evaluations},
	volume = {183},
	issn = {0964-1998, 1467-985X},
	url = {https://onlinelibrary.wiley.com/doi/10.1111/rssa.12522},
	doi = {10.1111/rssa.12522},
	language = {en},
	number = {2},
	urldate = {2021-07-24},
	journal = {Journal of the Royal Statistical Society: Series A (Statistics in Society)},
	author = {Gabrio, Andrea and Daniels, Michael and Baio, Gianluca},
	month = feb,
	year = {2020},
	pages = {607--629},
}

@article{gabrio_full_2019,
	title = {A full {Bayesian} model to handle structural ones and missingness in economic evaluations from individual-level data: {Handling} structural ones and missingness in economic evaluations},
	volume = {38},
	issn = {02776715},
	shorttitle = {A full {Bayesian} model to handle structural ones and missingness in economic evaluations from individual-level data},
	url = {https://onlinelibrary.wiley.com/doi/10.1002/sim.8045},
	doi = {10.1002/sim.8045},
	language = {en},
	number = {8},
	urldate = {2022-01-15},
	journal = {Statistics in Medicine},
	author = {Gabrio, Andrea and Mason, Alexina and Baio, Gianluca},
	month = apr,
	year = {2019},
	pages = {1399--1420},
}

@article{black_ce_1990,
	title = {The {CE} {Plane}: {A} {Graphic} {Representation} of {Cost}-{Effectiveness}},
	volume = {10},
	url = {https://doi.org/10.1177/0272989X9001000308},
	doi = {10.1177/0272989X9001000308},
	number = {3},
	journal = {Medical Decision Making},
	author = {Black, William C.},
	year = {1990},
	pages = {212--214},
}

@article{parent_handling_2013,
	title = {Handling {Item}-{Level} {Missing} {Data}: {Simpler} {Is} {Just} as {Good}},
	volume = {41},
	url = {https://doi.org/10.1177/0011000012445176},
	doi = {10.1177/0011000012445176},
	number = {4},
	journal = {The Counseling Psychologist},
	author = {Parent, Mike C.},
	year = {2013},
	pages = {568--600},
}

@article{vera_is_2021,
	title = {Is {Item} {Imputation} {Always} {Better}? {An} {Investigation} of {Wave}-{Missing} {Data} in {Growth} {Models}},
	volume = {28},
	issn = {1070-5511, 1532-8007},
	shorttitle = {Is {Item} {Imputation} {Always} {Better}?},
	url = {https://www.tandfonline.com/doi/full/10.1080/10705511.2020.1850289},
	doi = {10.1080/10705511.2020.1850289},
	language = {en},
	number = {4},
	urldate = {2023-10-26},
	journal = {Structural Equation Modeling: A Multidisciplinary Journal},
	author = {Vera, Juan Diego and Enders, Craig K.},
	month = jul,
	year = {2021},
	pages = {506--517},
}

@book{joint_formulary_committee_british_2023,
	title = {British national formulary},
	volume = {64},
	isbn = {978-0-85711-461-7},
	shorttitle = {{BNF} 86},
	url = {https://www.pharmaceuticalpress.com/product/british-national-formulary-bnf86-2/},
	language = {en},
	publisher = {Pharmaceutical Press},
	author = {Joint Formulary Committee},
	month = sep,
	year = {2023},
}

@article{rosel_what_2022,
	title = {What difference does multiple imputation make in longitudinal modeling of {EQ}-{5D}-{5L} data? {Empirical} analyses of simulated and observed missing data patterns},
	volume = {31},
	issn = {0962-9343, 1573-2649},
	shorttitle = {What difference does multiple imputation make in longitudinal modeling of {EQ}-{5D}-{5L} data?},
	url = {https://link.springer.com/10.1007/s11136-021-03037-3},
	doi = {10.1007/s11136-021-03037-3},
	language = {en},
	number = {5},
	urldate = {2023-10-26},
	journal = {Quality of Life Research},
	author = {Rösel, Inka and Serna-Higuita, Lina María and Al Sayah, Fatima and Buchholz, Maresa and Buchholz, Ines and Kohlmann, Thomas and Martus, Peter and Feng, You-Shan},
	month = may,
	year = {2022},
	pages = {1521--1532},
}

@article{fenwick_guide_2005,
	title = {A guide to cost-effectiveness acceptability curves},
	volume = {187},
	issn = {0007-1250, 1472-1465},
	url = {https://www.cambridge.org/core/product/identifier/S0007125000167133/type/journal_article},
	doi = {10.1192/bjp.187.2.106},
	language = {en},
	number = {2},
	urldate = {2023-10-20},
	journal = {British Journal of Psychiatry},
	author = {Fenwick, Elisabeth and Byford, Sarah},
	month = aug,
	year = {2005},
	pages = {106--108},
}

@article{fox-wasylyshyn_handling_2005,
	title = {Handling missing data in self-report measures},
	volume = {28},
	issn = {0160-6891, 1098-240X},
	url = {https://onlinelibrary.wiley.com/doi/10.1002/nur.20100},
	doi = {10.1002/nur.20100},
	language = {en},
	number = {6},
	urldate = {2023-09-30},
	journal = {Research in Nursing \& Health},
	author = {Fox-Wasylyshyn, Susan M. and El-Masri, Maher M.},
	month = dec,
	year = {2005},
	pages = {488--495},
}

@article{berchtold_treatment_2019,
	title = {Treatment and reporting of item-level missing data in social science research},
	volume = {22},
	issn = {1364-5579, 1464-5300},
	url = {https://www.tandfonline.com/doi/full/10.1080/13645579.2018.1563978},
	doi = {10.1080/13645579.2018.1563978},
	language = {en},
	number = {5},
	urldate = {2023-09-30},
	journal = {International Journal of Social Research Methodology},
	author = {Berchtold, André},
	month = sep,
	year = {2019},
	pages = {431--439},
}

@article{gelman_posterior_1996,
	title = {{POSTERIOR} {PREDICTIVE} {ASSESSMENT} {OF} {MODEL} {FITNESS} {VIA} {REALIZED} {DISCREPANCIES}},
	language = {en},
	number = {6},
	journal = {Statistica Sinica},
	author = {Gelman, Andrew and Meng, Xiao-Li and Stern, Hal},
	year = {1996},
	pages = {29},
}

@article{gabry_visualization_2019,
	title = {Visualization in {Bayesian} workflow},
	volume = {182},
	issn = {0964-1998, 1467-985X},
	url = {https://onlinelibrary.wiley.com/doi/10.1111/rssa.12378},
	doi = {10.1111/rssa.12378},
	language = {en},
	number = {2},
	urldate = {2022-08-25},
	journal = {Journal of the Royal Statistical Society: Series A (Statistics in Society)},
	author = {Gabry, Jonah and Simpson, Daniel and Vehtari, Aki and Betancourt, Michael and Gelman, Andrew},
	month = feb,
	year = {2019},
	pages = {389--402},
}

@article{li_random-effects_2007,
	title = {A random-effects {Markov} transition model for {Poisson}-distributed repeated measures with non-ignorable missing values},
	volume = {26},
	issn = {02776715, 10970258},
	url = {https://onlinelibrary.wiley.com/doi/10.1002/sim.2717},
	doi = {10.1002/sim.2717},
	language = {en},
	number = {12},
	urldate = {2023-07-06},
	journal = {Statistics in Medicine},
	author = {Li, Jinhui and Yang, Xiaowei and Wu, Yingnian and Shoptaw, Steven},
	month = may,
	year = {2007},
	pages = {2519--2532},
}

@article{lee_marginalized_2019,
	title = {Marginalized models for longitudinal count data},
	volume = {136},
	issn = {01679473},
	url = {https://linkinghub.elsevier.com/retrieve/pii/S0167947319300027},
	doi = {10.1016/j.csda.2019.01.001},
	language = {en},
	urldate = {2023-05-27},
	journal = {Computational Statistics \& Data Analysis},
	author = {Lee, Keunbaik and Joo, Yongsung},
	month = aug,
	year = {2019},
	pages = {47--58},
}

@article{jakobsen_when_2017,
	title = {When and how should multiple imputation be used for handling missing data in randomised clinical trials – a practical guide with flowcharts},
	volume = {17},
	issn = {1471-2288},
	url = {https://bmcmedresmethodol.biomedcentral.com/articles/10.1186/s12874-017-0442-1},
	doi = {10.1186/s12874-017-0442-1},
	language = {en},
	number = {1},
	urldate = {2023-05-12},
	journal = {BMC Medical Research Methodology},
	author = {Jakobsen, Janus Christian and Gluud, Christian and Wetterslev, Jørn and Winkel, Per},
	month = dec,
	year = {2017},
	pages = {162},
}

@book{fitzmaurice_longitudinal_2008,
	title = {Longitudinal data analysis},
	publisher = {CRC press},
	author = {Fitzmaurice, Garrett and Davidian, Marie and Verbeke, Geert and Molenberghs, Geert},
	year = {2008},
}

@misc{su_r2jags_2021,
	title = {R2jags: {Using} {R} to {Run} '{JAGS}'},
	copyright = {GPL ({\textgreater} 2)},
	shorttitle = {R2jags},
	url = {https://CRAN.R-project.org/package=R2jags},
	urldate = {2022-09-06},
	author = {Su, Yu-Sung and Yajima, Masanao},
	month = aug,
	year = {2021},
}

@article{plummer_jags_2003,
	title = {{JAGS}: {A} {Program} for {Analysis} of {Bayesian} {Graphical} {Models} using {Gibbs} {Sampling}},
	volume = {124},
	journal = {3rd International Workshop on Distributed Statistical Computing (DSC 2003); Vienna, Austria},
	author = {Plummer, Martyn},
	month = apr,
	year = {2003},
}

@article{harrington_randomised_2000,
	title = {Randomised comparison of the effectiveness and costs of community and hospital based mental health services for children with behavioural {disordersTopic}: 83;86},
	volume = {321},
	language = {en},
	journal = {BMJ},
	author = {Harrington, Richard and Peters, Sarah and Green, Jonathan and Byford, Sarah and Woods, Jane and McGowan, Ruth},
	year = {2000},
	pages = {5},
}

@book{curtis_unit_2019,
	series = {Unit {Costs} of {Health} and {Social} {Care}},
	title = {Unit {Costs} of {Health} and {Social} {Care} 2019},
	isbn = {978-1-911353-10-2},
	url = {https://kar.kent.ac.uk/id/eprint/79286},
	language = {en-GB},
	urldate = {2022-09-01},
	publisher = {PSSRU, University of Kent},
	author = {Curtis, Lesley A. and Burns, Amanda},
	month = dec,
	year = {2019},
}

@article{barrett_mental_2006,
	title = {Mental health provision for young offenders: service use and cost},
	volume = {188},
	issn = {0007-1250, 1472-1465},
	shorttitle = {Mental health provision for young offenders},
	url = {https://www.cambridge.org/core/product/identifier/S0007125000231917/type/journal_article},
	doi = {10.1192/bjp.bp.105.010108},
	language = {en},
	number = {6},
	urldate = {2022-04-27},
	journal = {British Journal of Psychiatry},
	author = {Barrett, Barbara and Byford, Sarah and Chitsabesan, Prathiba and Kenning, Cassandra},
	month = jun,
	year = {2006},
	pages = {541--546},
}

@article{furber_validity_2015,
	title = {The validity of the {Child} {Health} {Utility} instrument ({CHU9D}) as a routine outcome measure for use in child and adolescent mental health services},
	volume = {13},
	issn = {1477-7525},
	url = {http://hqlo.biomedcentral.com/articles/10.1186/s12955-015-0218-4},
	doi = {10.1186/s12955-015-0218-4},
	language = {en},
	number = {1},
	urldate = {2022-04-27},
	journal = {Health and Quality of Life Outcomes},
	author = {Furber, Gareth and Segal, Leonie},
	year = {2015},
	pages = {22},
}

@article{byford_cost-effectiveness_1999,
	title = {Cost-effectiveness analysis of a home-based social work intervention for children and adolescents who have deliberately poisoned themselves: {Results} of a randomised controlled trial},
	volume = {174},
	issn = {0007-1250, 1472-1465},
	shorttitle = {Cost-effectiveness analysis of a home-based social work intervention for children and adolescents who have deliberately poisoned themselves},
	url = {https://www.cambridge.org/core/product/identifier/S000712500015202X/type/journal_article},
	doi = {10.1192/bjp.174.1.56},
	language = {en},
	number = {1},
	urldate = {2022-09-01},
	journal = {British Journal of Psychiatry},
	author = {Byford, Sarah and Harrington, Richard and Torgerson, David and Kerfoot, Michael and Dyer, Elizabeth and Harrington, Val and Woodham, Adrine and Gill, Julia and McNiven, Faye},
	month = jan,
	year = {1999},
	pages = {56--62},
}

@article{mainzer_comparison_2021,
	title = {A comparison of multiple imputation strategies for handling missing data in multi‐item scales: {Guidance} for longitudinal studies},
	volume = {40},
	issn = {0277-6715, 1097-0258},
	shorttitle = {A comparison of multiple imputation strategies for handling missing data in multi‐item scales},
	url = {https://onlinelibrary.wiley.com/doi/10.1002/sim.9088},
	doi = {10.1002/sim.9088},
	language = {en},
	number = {21},
	urldate = {2021-08-22},
	journal = {Statistics in Medicine},
	author = {Mainzer, Rheanna and Apajee, Jemishabye and Nguyen, Cattram D. and Carlin, John B. and Lee, Katherine J.},
	month = sep,
	year = {2021},
	pages = {4660--4674},
}

@article{hollis_therapist-supported_2021,
	title = {Therapist-supported online remote behavioural intervention for tics in children and adolescents in {England} ({ORBIT}): a multicentre, parallel group, single-blind, randomised controlled trial},
	volume = {8},
	issn = {22150366},
	shorttitle = {Therapist-supported online remote behavioural intervention for tics in children and adolescents in {England} ({ORBIT})},
	url = {https://linkinghub.elsevier.com/retrieve/pii/S2215036621002352},
	doi = {10.1016/S2215-0366(21)00235-2},
	language = {en},
	number = {10},
	urldate = {2022-04-27},
	journal = {The Lancet Psychiatry},
	author = {Hollis, Chris and Hall, Charlotte L and Jones, Rebecca and Marston, Louise and Novere, Marie Le and Hunter, Rachael and Brown, Beverley J and Sanderson, Charlotte and Andrén, Per and Bennett, Sophie D and Chamberlain, Liam R and Davies, E Bethan and Evans, Amber and Kouzoupi, Natalia and McKenzie, Caitlin and Heyman, Isobel and Khan, Kareem and Kilgariff, Joseph and Glazebrook, Cristine and Mataix-Cols, David and Murphy, Tara and Serlachius, Eva and Murray, Elizabeth},
	month = oct,
	year = {2021},
	pages = {871--882},
}

@article{basu_regression_2012,
	title = {Regression {Estimators} for {Generic} {Health}-{Related} {Quality} of {Life} and {Quality}-{Adjusted} {Life} {Years}},
	volume = {32},
	issn = {0272-989X, 1552-681X},
	url = {http://journals.sagepub.com/doi/10.1177/0272989X11416988},
	doi = {10.1177/0272989X11416988},
	language = {en},
	number = {1},
	urldate = {2022-01-09},
	journal = {Medical Decision Making},
	author = {Basu, Anirban and Manca, Andrea},
	month = jan,
	year = {2012},
	pages = {56--69},
}

@article{sharples_amaze_2018,
	title = {Amaze: a double-blind, multicentre randomised controlled trial to investigate the clinical effectiveness and cost-effectiveness of adding an ablation device-based maze procedure as an adjunct to routine cardiac surgery for patients with pre-existing atrial fibrillation},
	volume = {22},
	issn = {1366-5278, 2046-4924},
	shorttitle = {Amaze},
	url = {https://www.journalslibrary.nihr.ac.uk/hta/hta22190},
	doi = {10.3310/hta22190},
	language = {en},
	number = {19},
	urldate = {2021-06-20},
	journal = {Health Technology Assessment},
	author = {Sharples, Linda and Everett, Colin and Singh, Jeshika and Mills, Christine and Spyt, Tom and Abu-Omar, Yasir and Fynn, Simon and Thorpe, Benjamin and Stoneman, Victoria and Goddard, Hester and Fox-Rushby, Julia and Nashef, Samer},
	month = apr,
	year = {2018},
	pages = {1--132},
}

\clearpage

\begin{table}
\caption{Distributional assumptions and link functions (in parentheses) for item-level models}
\label{table1}

\footnotesize
\renewcommand{\arraystretch}{1.12}
\setlength{\tabcolsep}{4pt}

\resizebox{\textwidth}{!}{%
\begin{tabular}{m{8cm}m{3.5cm}m{3cm}}
\hline
Selected cost categories: CAMHS, GP, and SENCO & Other cost components*   & Utility           \\ \hline
Cost: Normal (identity)  & Normal (identity)   & Normal (identity)  \\ 
Cost: Gamma  (log)       & Gamma (log)         & Beta (logit)       \\ 
Resource Use Count: Poisson with fixed effects (log) & Gamma (log)  & Beta (logit)     \\ 
Resource Use Count: Poisson with random effects (log) & Gamma (log)  & Beta (logit)    \\ 
\hline
\end{tabular}%
  }
\vspace{0.5mm}
\begin{flushleft}
CAMHS = Child and adolescent mental health; GP = General practice; SENCO = Special education need co-ordinator. * Other cost components include remaining costs, other medication use and intervention costs.
\end{flushleft}
\end{table}

\begin{table}[ht]
\centering
\caption{Mean and 95\% CIs of costs by treatment arm, incremental costs, QALYs by treatment arm, incremental QALYs and ICER of item-level models under MAR}
\label{table3}
\renewcommand{\arraystretch}{1.12}
\setlength{\tabcolsep}{4pt}
\footnotesize

\resizebox{\textwidth}{!}{%
\begin{tabular}{
  >{\raggedright\arraybackslash}p{2.5cm}
  S[table-format=3.0,group-separator={,},group-minimum-digits=3]
  S[table-format=3.0,group-separator={,},group-minimum-digits=3]
  S[table-format=2.0,group-separator={,},group-minimum-digits=3]
  S[table-format=1.3]
  S[table-format=1.3]
  S[table-format=1.4]
  S[table-format=5.0,group-separator={,},group-minimum-digits=4]
}
\toprule
{Model} & {Costs(Con)} & {Costs(Int)} & {Incremental Costs} &
{QALYs(Con)} & {QALYs(Int)} & {Incremental QALYs} & {ICER} \\
\midrule

\multirow[t]{2}{=}{Normal}
& 705 & 795 & 90 & 0.416 & 0.422 & 0.0058 & 15506  \\ 
& {(602, 807)} & {(673, 920)}  &  {(-70, 249)}  
& {(0.410, 0.422)} & {(0.416, 0.427)} & {(-0.0021, 0.0135)} & \\ 

\multirow[t]{2}{=}{Beta-Gamma} 
& 826 & 892 & 65 & 0.417 & 0.422 &  0.0051 & 12871 \\ 
& {(680, 981)} & {(731, 1060)} & {(-159, 299)}  
& {(0.410, 0.423)} & {(0.416, 0.427)} & {(-0.0036, 0.0141)} & \\

\multirow[t]{2}{=}{Poisson with fixed effects} 
& 806 & 880 & 74 & 0.416 & 0.420 & 0.0041 & 17953 \\ 
& {(678, 956)} & {(722, 1050)} & {(-165, 285)}  
& {(0.409, 0.422)} & {(0.413, 0.426)} & {(-0.0046, 0.0134)} & \\

\multirow[t]{2}{=}{Poisson with random effects}
& 795 & 882 & 87 & 0.416 & 0.421 & 0.0046 & 18981 \\ 
& {(659, 951)} & {(710, 1053)} & {(-166, 317)}  
& {(0.409, 0.423)} & {(0.414, 0.427)} & {(-0.0046, 0.0134)} & \\
\bottomrule
\end{tabular}%
  }

\vspace{0.5mm}
\begin{flushleft}
\footnotesize
Costs are measured using British pound (£). 95\% CIs are reported in parentheses. 
CI = Credible interval; Con = Control; ICER = Incremental cost-effectiveness ratio; 
Int = Intervention; MAR = Missing at random; QALY = Quality-adjusted life-year.
\end{flushleft}

\end{table}

\newpage  
\begin{appendices}
\section{Running Example Data Re-Cleaning Strategy} \label{appendix_strategy}
Given that the definition of ``missing items" in the original health economic analysis has a shaky ground, costs data in the ORBIT trial need to be re-cleaned. We employed a common definition of missing items --- i.e., data that are intended to be collected but not acquired at the time of analysis --- and recleaned the data based on the nature of cost questionnaire. 

The items in the ORBIT trial cost questionnaire can be broadly categorised into five main components: (1) specialist tics services, (2) community services, (3) inpatient services, (4) emergency services and (5) medication use. Each cost component can be further disaggregated into specific health and social care services. For instance, community services include general practitioners (GP), practice nurses, social workers, etc. For each cost component, the questionnaire starts by asking whether a certain pre-defined service has been used, followed by more detailed questions. Several sets of open-ended questions are also included to comprehensively collect information on services not covered by pre-defined items. A typical example is the ``other medication use'', as shown in Table~\ref{appendix:table1}. Apart from the three main medications, patients are required to report any unlisted medications and then provide detailed information about the dosage, daily frequency and duration of usage.

\begin{table}[ht]
\centering
\begin{tabular}{|P{43mm}|P{15mm}|P{10mm}|P{21mm}|P{24mm}|}
\hline
\rowcolor{lightgray} \multicolumn{5}{|c|}{Medication} \\
\hline
\rowcolor{lightgray} \multicolumn{1}{|c|}{Medication} & Yes/No & Dose & No. of times to be taken per day & No. of weeks dose taken in last 6 months \\
\hline
42.Clonidine &  &  &  & \\
\hline
43.Aripiprazole &  &  &  & \\
\hline
44.Risperidone &  &  &  & \\
\hline
45a.Other-please specify: &  &  &  & \\
\hline
45b.Other-please specify: &  &  &  & \\
\hline
45c.Other-please specify: &  &  &  & \\
\hline
45d.Other-please specify: &  &  &  & \\
\hline
45e.Other-please specify: &  &  &  & \\
\hline
45f.Other-please specify: &  &  &  & \\
\hline
\end{tabular}
\caption{Medication use section of the CA-SUS cost questionnaire}
\label{appendix:table1}
\end{table}

A set of criteria, tailored to our research objectives and not derived from published guidelines, has been established to identify items that should not be considered as missing: 
\begin{enumerate}
    \item If the patient indicates no use of the service, and the following resource-use questions for the same service are missing, then the remaining questions are not missing and should be assigned as zero resource use;
    \item If the questions are answered in the opposite way --- the patient does not report whether the service has been used but reports zeros in the remaining questions for the same service to indicate unused resource items, then the patient did not use the service;
    \item In the case where a few sets of open-ended questions have been provided to collect unlisted resource use, if the first set of the open-ended questions are completely observed but the remaining sets of open-ended questions are all unobserved, they are not missing and replaced by zeros.
\end{enumerate}

\newpage
\section{Proportion of Missingness} 
\label{appendix_proportion}

\begin{table}[h!] 
\centering
\renewcommand{\arraystretch}{1.2}
\begin{threeparttable}
\begin{tabular}{|P{42mm}
|S[table-format=3.0, table-column-width=4mm] >{\raggedright\arraybackslash} P{15mm}
|S[table-format=3.0, table-column-width=4mm] >{\raggedright\arraybackslash} P{15mm}
|S[table-format=3.0, table-column-width=4mm] >{\raggedright\arraybackslash} P{15mm}|}
\hline
\multicolumn{7}{|l|}{\textit{Missingness}}\\
\hline
 & \multicolumn{2}{c|}{Intervention} & \multicolumn{2}{c|}{Control}   & \multicolumn{2}{c|}{Total}     \\
 & \multicolumn{2}{c|}{(N = 112)}    & \multicolumn{2}{c|}{(N = 112)} & \multicolumn{2}{c|}{(N = 224)} \\
\hline
Intervention costs & 1 & (0.9\%) & 1 & (0.9\%) & 2 & (0.9\%) \\
\hline
\multicolumn{7}{|l|}{Heatlh care costs}\\
\hline
\hspace{1mm} Baseline & 56 & (50.0\%) & 60 & (53.6\%) & 116 & (51.8\%) \\
\hline
\hspace{2mm} Item-level missingness   & 56 & (50.0\%) & 59 & (52.7\%) & 115 & (51.3\%) \\
\hline
\hspace{2mm} Complete missingness & 0 & (0.0\%) & 1 & (0.9\%) & 1 & (0.4\%) \\
\hline
\hspace{1mm} 3 months & 66 & (58.9\%) & 65 & (58.0\%) & 131 & (58.5\%) \\
\hline
\hspace{2mm} Item-level missingness   & 55 & (49.1\%) & 53 & (47.3\%) & 108 & (48.2\%) \\
\hline
\hspace{2mm} Complete missingness & 11 & (9.8\%) & 12 & (10.7\%) & 33 & (14.7\%) \\
\hline
\hspace{1mm} 6 months & 76 & (67.9\%) & 77 & (68.8\%) & 153 & (68.3\%) \\
\hline
\hspace{2mm} Item-level missingness   & 56 & (50.0\%) & 58 & (51.8\%) & 114 & (50.9\%) \\
\hline
\hspace{2mm} Complete missingness & 20 & (17.9\%) & 19 & (17.0\%) & 39 & (17.4\%) \\
\hline
\multicolumn{7}{|l|}{CHU-9D}\\
\hline
\hspace{2mm} Baseline &  0 & (0.0\%)  &  0 & (0.0\%)  &  0 & (0.0\%) \\
\hline
\hspace{2mm} 3 months & 23 & (20.5\%) & 20 & (17.9\%) & 43 & (19.2\%) \\
\hline
\hspace{2mm} 6 months & 48 & (42.9\%) & 36 & (32.1\%) & 84 & (37.5\%) \\
\hline
\multicolumn{7}{|l|}{\textit{Complete cases}}\\
\hline
Complete total costs & 30 & (26.8\%) & 30 & (26.8\%) & 60 & (26.8\%) \\
\hline
Complete QALYs & 63 & (56.3\%) & 74 & (66.1\%) & 137 & (61.2\%) \\
\hline
Complete cases & 17 & (15.2\%) & 18 & (16.1\%) & 35 & (15.6\%) \\
\hline
\end{tabular}

\caption{Number and proportion of patients with missing data at each time point and complete cases by treatment arm.}
\label{appendix:table3}

\begin{tablenotes}[flushleft]
\item CHU-9D = Child Health Utility 9D; QALYs = Quality-adjusted life-years. Complete missingness refers to missingness caused by completely missing questionnaires while item-level missingness means missingness due to missing items.
\end{tablenotes}
\end{threeparttable}
\end{table}

\section{Cost Item Identification for Item-Level Modelling} \label{appendix_items}
A strategy is proposed to streamline the number of items in the cost questionnaire to a quantity that is practical to handle. The strategy reduces the scope of the questionnaire from 480 to 132 items, which effectively provides a basis for the key cost items to be selected from. The strategy requires the exclusion of questions that are: (1) beyond the scope of health sector perspective because the health sector perspective is the perspective adopted by the primary CEA in the ORBIT trial and can serve as a reference for our model to be compared to; (2) consistently completed with blank or zero values or missing data for all participants at every time point because these questions, although being included in the questionnaire, are in fact not utilised in the analysis; and (3) not incorporated in the original health economic analysis for all participants for the consistency purpose.

However, 132 items are still too many to model. It is still crucial to identify suitable cost items that can be modelled at item level. Ideally, they should be chosen under the following criteria which were specifically designed for the case study: 

\begin{enumerate}
    \item In each cost item, the amount of missingness should not be negligible over time. In the opposite of the former case, previous guideline on missing data in clinical trials has confirmed that missing values can be disregarded when the proportion of missingness is less than 5\% as a rule of thumb \parencite{jakobsen_when_2017}. However, this rule should not be the sole missingness-based criterion for selecting cost items to model. Instead, decisions should consider both the proportion of missingness within individual cost items and the cumulative impact across the dataset.
    \item The proportion of 0 among observed values within each selected cost item should remain at a relatively small level. If a substantial proportion of observed values in one specific cost item is 0, for instance, over 90\% or the extreme case of 100\%, substituting missing values by 0s may be a reasonable approach. In such extreme cases, item-level modelling is generally not feasible, making replacing these missing items with 0s the only practical option. However, for cost items with a lower proportion of observed 0s, item-level modelling should be considered instead.
    \item When considering whether cost items could be modelled as count data, observations within each selected item should be consistent. A cost category is defined to be consistent if it is collected in a similar manner and presents same data type across all observations at the three time points. A typical example of inconsistent cost item that may not be suitable for modelling as count data at item level is dosage of other medication use because it is rarely reported in the same unit for the whole sample.
\end{enumerate}

Each cost component has been checked in line with the criteria (Table~\ref{appendix:table2}). Other medication use exhibits a higher degree of missingness compared to other cost categories, thereby becoming the main driver of missing cost data at each time point. However, modelling other medication use at the item level is too complicated to implement because the data are inconsistent, with people reporting their medication dosages using different scales and units of measure. 

An attempt has been made to convert the unit of other medication use to a consistent format based on the British National Formulary (BNF) \parencite{joint_formulary_committee_british_2023}, the pharmaceutical reference book provided by NICE in the UK. Unfortunately the BNF does not have the recommended dose for all reported medications. Therefore, additional assumptions are required, which can add extra complexities to the original health economic analysis and deviate from the purpose of this work. To ease model implementation and without affecting the aim of the analysis, the costs of other medication use have been calculated based on the relevant items and will be modelled as continuous data. 

\fontsize{9}{10} \selectfont
\begin{longtable}{|p{1.5cm}|p{1cm}|p{1cm}|p{1.2cm}|p{1cm}|p{1cm}|p{1.2cm}|p{1cm}|p{1cm}|p{1.2cm}|}
\caption{Cost item identification for item-level modelling} \label{appendix:table2} \\ \hline
& \multicolumn{3}{c|}{Baseline}  & \multicolumn{3}{c|}{3 months} & \multicolumn{3}{c|}{6 months} \\ \hline
& Missing (\%) & Zeros among observed values (\%) & Consistent scale & Missing (\%) & Zeros among observed values (\%) & Consistent scale & Missing (\%) & Zeros among observed values (\%) & Consistent scale  \\ \hline
\endfirsthead 

\hline
& \multicolumn{3}{c|}{Baseline}  & \multicolumn{3}{c|}{3 months} & \multicolumn{3}{c|}{6 months} \\ \hline
& Missing (\%) & Zeros among observed values (\%) & Consistent scale & Missing (\%) & Zeros among observed values (\%) & Consistent scale & Missing (\%) & Zeros among observed values (\%) & Consistent scale  \\ \hline
\endhead 

\multicolumn{10}{r}{{Continued on next page}} \\
\endfoot 

\hline
\endlastfoot 

\multicolumn{10}{|l|}{\textbf{Specialist service}} \\ \hline
\multicolumn{10}{|l|}{Tic disorder clinic} \\ \hline 
at clinic & 0.4 &  94.6 & Yes & 10.3 &  95.5 & Yes & 17.4 &  96.8 & Yes \\ \hline
at home   & 0.4 & 100.0 & Yes & 10.3 & 100.0 & Yes & 17.4 & 100.0 & Yes \\ \hline
by phone  & 0.4 &  99.1 & Yes & 10.3 & 100.0 & Yes & 17.4 & 100.0 & Yes \\ \hline
if OOP    & 0.4 & 100.0 & Yes & 10.3 & 98.0 & Yes & 17.4 & 100.0 & Yes \\ \hline
\multicolumn{10}{|l|}{CAMHS services} \\ \hline
at clinic  & 0.4 & 75.3 & Yes & 10.3  &  79.6 & Yes & 17.4 & 78.4 & Yes \\ \hline
at home    & 0.4 & 99.6 & Yes & 10.3  & 100.0 & Yes & 17.4 & 99.5 & Yes \\ \hline
by phone   & 0.4 & 94.6 & Yes & 10.3  &  98.0 & Yes & 17.4 & 96.2 & Yes \\ \hline
if OOP     & 0.4 & 99.6 & Yes & 10.3 & 100.0 & Yes & 17.4 & 100.0 & Yes \\ \hline
\multicolumn{10}{|l|}{Paediatrician in hospital} \\ \hline
at clinic  & 0.4 &  85.2 & Yes & 10.3 &  91.5 & Yes & 17.4 &  89.2  & Yes \\ \hline
at home    & 0.4 & 100.0 & Yes & 10.3 & 100.0 & Yes & 17.4 & 100.0 & Yes \\ \hline
by phone   & 0.4 & 100.0 & Yes & 10.3 &  99.5 & Yes & 17.4 & 100.0 & Yes \\ \hline
if OOP     & 0.4 &  99.1 & Yes & 10.3 &  99.5 & Yes & 17.4 & 100.0 & Yes \\ \hline
\multicolumn{10}{|l|}{Paediatrician in community} \\ \hline
at clinic & 0.4  &  85.2 & Yes & 10.3  &  96.0 & Yes & 17.4 &  89.2 & Yes \\ \hline
at home   & 0.4  & 100.0 & Yes & 10.3  & 100.0 & Yes & 17.4 & 100.0 & Yes \\ \hline
by phone  & 0.4  & 100.0 & Yes & 10.3  & 100.0 & Yes & 17.4 & 100.0 & Yes \\ \hline
if OOP    & 0.4  &  99.1 & Yes & 10.3  &  99.5 & Yes & 17.4 & 100.0 & Yes \\ \hline
\multicolumn{10}{|l|}{Psychiatrist} \\ \hline
at clinic & 0.4 &  97.3 & Yes & 10.3 &  97.0 & Yes & 17.4 &  96.8 & Yes\\ \hline
at home   & 0.4 & 100.0 & Yes & 10.3 & 100.0 & Yes & 17.4 & 100.0 & Yes\\ \hline
by phone  & 0.4 & 100.0 & Yes & 10.3 & 100.0 & Yes & 17.4 &  99.5 & Yes\\ \hline
if OOP    & 0.4 &  99.6 & Yes & 10.3 &  99.0 & Yes & 17.4 & 100.0 & Yes\\ \hline
\multicolumn{10}{|l|}{Neurologist}  \\ \hline
at clinic & 0.4 &  97.3 & Yes & 10.3 &  98.5 & Yes & 17.4 &  94.6 & Yes\\ \hline
at home   & 0.4 & 100.0 & Yes & 10.3 & 100.0 & Yes & 17.4 & 100.0 & Yes\\ \hline
by phone  & 0.4 & 100.0 & Yes & 10.3 & 100.0 & Yes & 17.4 & 100.0 & Yes\\ \hline
if OOP    & 0.4 & 100.0 & Yes & 10.3 & 100.0 & Yes & 17.4 & 100.0 & Yes\\ \hline
\multicolumn{10}{|l|}{Psychologist} \\ \hline
at clinic & 0.4 &  99.1 & Yes & 10.3 &  99.0 & Yes & 17.4 &  97.8 & Yes\\ \hline
at home   & 0.4 & 100.0 & Yes & 10.3 & 100.0 & Yes & 17.4 & 100.0 & Yes\\ \hline
by phone  & 0.4 &  99.6 & Yes & 10.3 & 100.0 & Yes & 17.4 & 100.0 & Yes\\ \hline
if OOP    & 0.4 &  99.6 & Yes & 10.3 & 100.0 & Yes & 17.4 & 100.0 & Yes\\ \hline
\multicolumn{10}{|l|}{Speech and language therapist}  \\ \hline
at clinic & 0.4 &  97.8 & Yes & 10.3 & 96.0  & Yes & 17.4 &  98.4 & Yes\\ \hline
at home   & 0.4 & 100.0 & Yes & 10.3 & 100.0 & Yes & 17.4 & 100.0 & Yes\\ \hline
by phone  & 0.4 & 100.0 & Yes & 10.3 & 100.0 & Yes & 17.4 & 100.0 & Yes\\ \hline
if OOP    & 0.4 &  99.6 & Yes & 10.3 &  99.5 & Yes & 17.4 & 100.0 & Yes \\ \hline
\multicolumn{10}{|l|}{Occupational therapist}  \\ \hline
at clinic & 0.4 &  99.1 & Yes & 10.3 &  97.0 & Yes & 17.4 &  97.8 & Yes \\ \hline
at home   & 0.4 &  99.6 & Yes & 10.3 &  99.5 & Yes & 17.4 &  99.5 & Yes \\ \hline
by phone  & 0.4 &  99.6 & Yes & 10.3 & 100.0 & Yes & 17.4 & 100.0 & Yes \\ \hline
if OOP    & 0.4 & 100.0 & Yes & 10.3 &  99.5 & Yes & 17.4 & 100.0 & Yes \\ \hline
\multicolumn{10}{|l|}{Other specialist services} \\ \hline
description  & 0.4 &  96.9 & Yes& 10.3 &  95.0 & Yes & 17.4 &  94.6 & Yes \\ \hline
at clinic    & 0.4 &  97.3 & Yes& 10.3 &  96.0 & Yes & 17.4 &  95.1 & Yes \\ \hline
at home      & 0.4 &  99.6 & Yes& 10.3 &  99.5 & Yes & 17.4 &  99.5 & Yes \\ \hline
by phone     & 0.4 & 100.0 & Yes& 10.3 & 100.0 & Yes & 17.4 & 100.0 & Yes \\ \hline
if OOP       & 0.4 & 100.0 & Yes& 10.3 &  99.5 & Yes & 17.4 & 100.0 & Yes \\ \hline
\multicolumn{10}{|l|}{\textbf{Community service}} \\ \hline
\multicolumn{10}{|l|}{GP} \\ \hline
at clinic    & 0.4 &  71.3 & Yes & 10.3 & 79.6 & Yes & 17.4 &  78.4 & Yes \\ \hline
at home      & 0.4 & 100.0 & Yes & 10.3 & 99.5 & Yes & 17.4 & 100.0 & Yes \\ \hline
by phone     & 0.4 &  99.1 & Yes & 10.3 & 98.5 & Yes & 17.4 &  98.4 & Yes \\ \hline
\multicolumn{10}{|l|}{Practice Nurse}  \\ \hline
at clinic    & 0.4 &  91.0 & Yes & 10.3 &  95.0 & Yes & 17.4 &  96.2 & Yes \\ \hline
at home      & 0.4 & 100.0 & Yes & 10.3 & 100.0 & Yes & 17.4 & 100.0 & Yes \\ \hline
by phone     & 0.4 & 100.0 & Yes & 10.3 &  99.5 & Yes & 17.4 & 100.0 & Yes \\ \hline
\multicolumn{10}{|l|}{Social Worker}  \\ \hline
at clinic    & 0.4 & 98.7 & Yes & 10.3 &  99.5 & Yes & 17.4 &  98.9 & Yes \\ \hline
at home      & 0.4 & 98.7 & Yes & 10.3 &  99.0 & Yes & 17.4 &  98.9 & Yes \\ \hline
by phone     & 0.4 & 99.1 & Yes & 10.3 & 100.0 & Yes & 17.4 & 100.0 & Yes \\ \hline
\multicolumn{10}{|l|}{SENCO}  \\ \hline
no. contact  & 0.4 &  72.2 & Yes & 10.3 &  76.6 & Yes & 17.4 &  81.1 & Yes \\ \hline
if OOP       & 0.4 & 100.0 & Yes & 10.3 & 100.0 & Yes & 17.4 & 100.0 & Yes \\ \hline
\multicolumn{10}{|l|}{Educational Psychologist} \\ \hline
at clinic    & 0.4 &  96.9 & Yes & 10.3 &  97.0 & Yes & 17.4 &  93.5 & Yes\\ \hline
at home      & 0.4 & 100.0 & Yes & 10.3 & 100.0 & Yes & 17.4 & 100.0 & Yes\\ \hline
by phone     & 0.4 & 100.0 & Yes & 10.3 & 100.0 & Yes & 17.4 &  99.5 & Yes\\ \hline
if OOP       & 0.4 &  99.6 & Yes & 10.3 & 100.0 & Yes & 17.4 & 100.0 & Yes\\ \hline
\multicolumn{10}{|l|}{Any parent group} \\ \hline
description  & 0.4 &  93.7 & Yes & 10.3 &  96.5 & Yes & 17.4 &  95.7 & Yes \\ \hline
no. contact  & 0.4 &  95.1 & Yes & 10.3 &  97.0 & Yes & 17.4 &  96.8 & Yes \\ \hline
if OOP       & 0.4 & 100.0 & Yes & 10.3 & 100.0 & Yes & 17.4 & 100.0 & Yes \\ \hline
\multicolumn{10}{|l|}{Play therapist}  \\ \hline
no. contact  & 0.4 & 100.0 & Yes & 10.3 & 99.5 & Yes & 17.4 &  99.5 & Yes \\ \hline
if OOP       & 0.4 & 100.0 & Yes & 10.3 & 99.5 & Yes & 17.4 & 100.0 & Yes \\ \hline
\multicolumn{10}{|l|}{Art/drama/music therapist} \\ \hline
no. contact  & 0.4 & 99.1 & Yes & 10.3 &  99.5 & Yes & 17.4 &  98.9 & Yes\\ \hline
if OOP       & 0.4 & 99.6 & Yes & 10.3 & 100.0 & Yes & 17.4 & 100.0 & Yes\\ \hline
\multicolumn{10}{|l|}{Physio}  \\ \hline
no. contact  & 0.4 & 98.2 & Yes & 10.3 & 98.0 & Yes & 17.4 & 96.2 & Yes\\ \hline
if OOP       & 0.4 & 99.6 & Yes & 10.3 & 98.5 & Yes & 17.4 & 99.5 & Yes\\ \hline
\multicolumn{10}{|l|}{Diet/nutricianist} \\ \hline
no. contact  & 0.4 &  96.4 & Yes & 10.3 &  98.5 & Yes & 17.4 & 100.0 & Yes\\ \hline
if OOP       & 0.4 & 100.0 & Yes & 10.3 & 100.0 & Yes & 17.4 & 100.0 & Yes\\ \hline
\multicolumn{10}{|l|}{Osteopath} \\ \hline
no. contact  & 0.4 & 98.7 & Yes & 10.3 & 100.0 & Yes & 17.4 & 98.9 & Yes\\ \hline
if OOP       & 0.4 & 99.6 & Yes & 10.3 & 100.0 & Yes & 17.4 & 98.9 & Yes\\ \hline
\multicolumn{10}{|l|}{Alternative therapy} \\ \hline
description  & 0.4 & 96.4 & Yes & 10.3 & 97.0 & Yes & 17.4 & 97.3 & Yes\\ \hline
no. contact  & 0.4 & 96.4 & Yes & 10.3 & 97.5 & Yes & 17.4 & 97.3 & Yes\\ \hline
if OOP       & 0.4 & 96.9 & Yes & 10.3 & 97.5 & Yes & 17.4 & 97.8 & Yes\\ \hline
\multicolumn{10}{|l|}{Specialist orthodontics} \\ \hline
no. contact  & 0.4  & 91.0 & Yes & 10.3 & 89.6 & Yes & 17.4 & 91.4 & Yes\\ \hline
if OOP       & 0.4  & 97.8 & Yes & 10.3 & 98.0 & Yes & 17.4 & 98.9 & Yes\\ \hline
\multicolumn{10}{|l|}{Any other professionals} \\ \hline
description  & 0.4  & 90.1 & Yes & 10.3 &  87.1 & Yes & 17.4 &  88.1 & Yes \\ \hline
at clinic    & 0.4 &  90.6 & Yes & 10.3 &  89.1 & Yes & 17.4 &  89.2 & Yes \\ \hline
at home      & 0.4 &  99.6 & Yes & 10.3 & 100.0 & Yes & 17.4 &  99.5 & Yes \\ \hline
by phone     & 0.4 & 100.0 & Yes & 10.3 & 100.0 & Yes & 17.4 & 100.0 & Yes \\ \hline
if OOP       & 0.4 & 100.0 & Yes & 10.3 &  98.0 & Yes & 17.4 & 100.0 & Yes \\ \hline
\multicolumn{10}{|l|}{\textbf{Hospital services}}  \\ \hline
\multicolumn{10}{|l|}{Inpatient} \\ \hline
reason for stay in hospital & 0.4  & 98.2 & Yes & 10.3 & 99.0 & Yes & 17.4 & 98.9 & Yes \\ \hline
no. nights    & 0.4 & 98.2 & Yes & 10.3 & 99.0 & Yes & 17.4 & 99.5 & Yes \\ \hline
\multicolumn{10}{|l|}{A and E attendance} \\ \hline
reason for stay in hospital & 0.4 & 92.4 & Yes & 10.3 & 95.0 & Yes & 17.4 & 94.6 & Yes \\ \hline
no. nights    & 0.4 & 92.4 & Yes & 10.3 & 95.0 & Yes & 17.4 & 94.6 & Yes \\ \hline
\multicolumn{10}{|l|}{\textbf{Medication}} \\ \hline
\multicolumn{10}{|l|}{Clonidine} \\ \hline
does          & 0.4 & 94.2 & Yes & 10.3 & 93.5 & Yes & 17.4 & 94.1 & Yes \\ \hline
times per day & 0.4 & 94.2 & Yes & 10.3 & 93.5 & Yes & 17.4 & 94.1 & Yes \\ \hline
weeks taken   & 0.4 & 94.2 & Yes & 10.3 & 93.5 & Yes & 17.4 & 94.1 & Yes \\ \hline
\multicolumn{10}{|l|}{Aripiprazole} \\ \hline
does          & 0.4 & 96.4 & Yes & 10.3 & 97.0 & Yes & 17.4 & 97.3 & Yes \\ \hline
times per day & 0.4 & 96.4 & Yes & 10.3 & 97.0 & Yes & 17.4 & 97.3 & Yes \\ \hline
weeks taken   & 0.4 & 96.4 & Yes & 10.3 & 97.0 & Yes & 17.4 & 97.3 & Yes \\ \hline
\multicolumn{10}{|l|}{Risperidone} \\ \hline
does          & 0.4 & 98.7 & Yes & 10.3 & 99.0 & Yes & 17.4 & 99.5 & Yes \\ \hline
times per day & 0.4 & 98.7 & Yes & 10.3 & 99.0 & Yes & 17.4 & 99.5 & Yes \\ \hline
weeks taken   & 0.4 & 98.7 & Yes & 10.3 & 99.0 & Yes & 17.4 & 99.5 & Yes \\ \hline
\multicolumn{10}{|l|}{Other medication}  \\ \hline 
description   & 51.8 & 36.1 & No  & 58.5 & 29.0 & No  & 68.3 & 15.5 & No  \\ \hline 
does          & 51.8 & 38.0 & No  & 58.5 & 32.3 & No  & 68.3 & 22.5 & No  \\ \hline
times per day & 51.8 & 38.9 & Yes & 58.5 & 30.1 & Yes & 68.3 & 15.5 & Yes \\ \hline
weeks taken   & 51.8 & 38.0 & Yes & 58.5 & 30.1 & Yes & 68.3 & 16.9 & Yes \\ 
\end{longtable}
\noindent
CAMHS = Child and adolescent mental health; OOP = Out-of-pocket; SENCO = Special education need co-ordinator.
\normalsize

\section{Transition Model Formulation} 
\label{appendix_transition}
Let $\bm{\bm{Y}_{i}} = (Y_{i1}, \ldots, Y_{iJ})$ denote any general responses over time $J$ for individual $i$. The joint distribution of responses can be expressed as a series of conditional distributions given a set of covariates $\bm{X}_{i}$ \parencite{fitzmaurice_longitudinal_2008}: 
\begin{equation*}
    p(Y_{i1}, \ldots, Y_{iJ} \mid \bm{X}_{i}) = \prod^J_{j=1}p(Y_{ij} | Y_{i1}, \ldots, Y_{i(j-1)} ; \bm{X}_{i})
\end{equation*}
By doing this, the model can effectively account for the serial dependence between longitudinal responses and focus on the mean regression structure \parencite{heagerty_marginalized_2000}, which aligns with the interest of CEAs.

The method models the conditional distribution of current response as a function of covariates and past responses. Suppose the response $Y_{ij}$ for patient $i$ at current time point $j$ depends on the history of past responses up to the previous time point $j-1$, represented by $H_{ij} = (Y_{i1}, \ldots, Y_{i(j-1)})$. The generalised linear model is then specified as \parencite{zeger_markov_1988}:
\begin{equation*} 
    h^{-1} \{\text{E}(Y_{ij} | \bm{X}_{i}, H_{ij})\} = \bm{X}_{i} \bm{\beta} + \sum_{r=1}^R \alpha_r f_r(H_{ij})
\end{equation*}
where $h^{-1}(.)$ is the link function, $f_r(.)$ denotes a function of the history of past responses at lag $r$ under an $R$th-order Markov assumption, and $\bm{\beta}$ and $\alpha_r$ are coefficients describing the effects of covariates and past responses on the current outcome, respectively.

Under a \textit{first-order Markov assumption}, i.e.~response $\bm{Y}_{ij}$ for patient $i$ at current time point $j$ only depends on response $\bm{Y}_{i(j-1)}$ at immediate previous time point $j-1$, the generalised linear model now takes the form of:
\begin{equation*} 
    h^{-1} \{\text{E}(\bm{Y}_{ij} | \bm{X}_{i}, \bm{Y}_{i(j-1)})\} = \bm{X}_{i} \bm{\beta} + \alpha f(\bm{Y}_{i(j-1)})
\end{equation*}
It is theoretically plausible to consider a higher-order Markov process. However, second- or higher-order Markov models have not been used for the analysis of our motivating example due to practical reasons: first, there are only two follow-up points after baseline data collection in our case study; second, computational challenges could arise when applying higher-order Markov processes to data with a relatively limited sample size.

\section{Identification of Missingness Predictors} \label{appendix_predictors}
The original health economic analysis plan of the trial identifies site as the only missingness predictor and includes it as a covariate in its regression models, together with costs and utilities at baseline. However, since the data have been re-cleaned following our previous outlined strategy, and considering the item-level model is a completely different model from the one utilised by the original analysis (i.e.~the model at the most aggregated level of total costs and total QALYs), new missingness predictors have been explored for item-level imputation.

In the new dataset, missingness indicators have been created for utility values and the item-level cost outcomes at all three time points. Then logistic regression has been performed to find out demographic variables associated with the missing outcomes at each time point. At baseline, given that there are no missing utility values and only one missing observation for most cost categories, logistic regression has been conducted only for the other medication use.

\begin{table}
\footnotesize
\renewcommand{\arraystretch}{1.5}
\begin{tabular}{p{3.0cm} >{\raggedleft\arraybackslash}p{0.5cm} p{2.0cm} p{0.05cm}
                            >{\raggedleft\arraybackslash}p{0.7cm} p{2.0cm} p{0.05cm}
                            >{\raggedleft\arraybackslash}p{0.5cm} p{2.0cm} p{0.05cm}}
\hline
& \multicolumn{3}{p{2.55cm}}{\centering CAMHS, GP, SENCO, Remaining costs} & \multicolumn{3}{p{2.75cm}}{\centering Other medication use}     & \multicolumn{3}{p{2.55cm}}{\centering Utility}       \\ \hline
\multicolumn{7}{l}{\textit{Baseline}}                                                     \\ 
Age                  & & &               & 1.02   & (0.89, 1.18) & & &                    \\ 
Gender               & & &               & 1.25   & (0.62, 2.55) & & &                    \\ 
Site                 & & &               & 2.86***& (1.64, 5.07) & & &                    \\ 
Comorbidity:ADHD     & & &               & 0.38*  & (0.17, 0.81) & & &                    \\ 
Comorbidity:OCD      & & &               & 0.26** & (0.10, 0.64) & & &                    \\ 
Treatment arm        & & &               & 0.84   & (0.48, 1.47) & & &                    \\ 
\multicolumn{7}{l}{\textit{3 months}}                                                     \\ 
Age                  & 1.10  & (0.88, 1.36) && 0.96 & (0.83, 1.10)    && 1.20* & (1.02, 1.43)  \\  
Gender               & 0.53  & (0.12, 1.72) && 0.81 & (0.40, 1.62)    && 0.75  & (0.61, 2.46)  \\ 
Site                 & 0.35* & (0.12, 0.90) && 2.58*** & (1.48, 4.55) && 0.87  & (0.43, 1.72)  \\ 
Comorbidity:ADHD     & 1.86  & (0.61, 5.16) && 0.49 & (0.23, 1.02)    && 2.54* & (1.09, 5.74)  \\
Comorbidity:OCD      & 1.23  & (0.27, 4.25) && 0.46 & (0.19, 1.08)    && 1.06  & (0.33, 2.92)  \\ 
Treatment arm        & 0.94  & (0.38, 2.28) && 1.01 & (0.58, 1.76)    && 1.29  & (0.65, 2.58)  \\ 
\multicolumn{7}{l}{\textit{6 months}}                                                      \\ 
Age                  & 1.03  & (0.86, 1.23) && 1.00   & (0.86, 1.16)  && 1.17* & (1.01, 1.35)  \\ 
Gender               & 0.67  & (0.23, 1.65) && 0.65   & (0.32, 1.35)  && 1.24  & (0.61, 2.46)  \\ 
Site                 & 0.62* & (0.30, 1.26) && 2.63** & (1.45, 4.89)  && 0.64  & (0.36, 1.12)  \\ 
Comorbidity:ADHD     & 2.00  & (0.82, 4.61) && 0.81   & (0.37, 1.83)  && 1.69  & (0.80, 3.57)  \\ 
Comorbidity:OCD      & 1.27  & (0.39, 3.50) && 0.29** & (0.12, 0.70)  && 1.41  & (0.59, 3.30)  \\ 
Treatment arm        & 1.08  & (0.53, 2.19) && 0.93   & (0.51, 1.68)  && 1.71  & (0.98, 3.03)  \\ \hline
\end{tabular}
\caption{Odds ratios (95\% confidence intervals) from logistic regression for missingness indicator of outcomes}
\label{appendix:table4}
{\small ADHD = Attention deficit hyperactivity disorder; CAMHS = child and adolescent mental health; GP = General practice; OCD = Obsessive compulsive disorder; SENCO = Special education need co-ordinator. \\
* indicates a statistical significance at 0.05; \\
** indicates a statistical significance at 0.01; \\
*** indicates a statistical significance at 0.001. \\}
\end{table}

Logistic regression has revealed associations between certain demographic variables and missingness in outcomes, as shown in Table~\ref{appendix:table3}. Site has been found to be associated with missing costs for all cost components; comorbidity is associated with missingness in other medication use at baseline and 6 months, and with missing utility values at 3 months; age is associated with missing utilities at 3 and 6 months. Based on these findings, our initial imputation model includes the site as a predictor for missing CAMHS, GP, SENCO and remaining costs, while site and comorbidity are included in the model for other medication use, and age and comorbidity in the utility model. 

Throughout this work, the missingness mechanism is assumed to be MAR, as it is the recommended starting point for missing data analysis in many cases \parencite{faria_guide_2014, carpenter_missing_2021}.

\section{Sensitivity Analysis for Poisson Models} \label{appendix_poisson}
The $\theta = 0.5$ in Poisson models is determined via a sensitivity analysis over a set of theoretically plausible values: 0.1, 0.2, ..., 1 (see Table~\ref{appendix:table4}). Higher values than 0.5 are excluded from the sensitivity analysis to ensure the data after transformation, although only serving as covariates, are distinguishable from actual observed values of 1, and to prevent potential overestimation of cost estimates, particularly in the Poisson model with random effects. Among the remaining values, $\theta = 0.5$ results in the lowest DIC across both the Poisson model with fixed effects and the Poisson model with random effects. Consequently, $\theta = 0.5$ is chosen as it provides a balance between achieving a better model fit and avoiding overestimation in cost estimates across both models.

\begin{sidewaystable}
\caption{Sensitivity analysis for the choice of $\theta$ in Poisson models} 
\label{appendix:table5}
\renewcommand{\arraystretch}{1.5}
\footnotesize
\begin{adjustbox}{max width=\textheight}
\begin{tabular}{p{1cm}p{1cm}p{3cm}p{3cm}p{3.5cm}p{3cm}p{3cm}p{3.5cm}}
\hline
$\theta$ & DIC* & Costs(Con) & Costs(Int) & Incremental Costs & QALYs(Con) & QALYs(Int) & Incremental QALYs \\ \hline
\multicolumn{8}{l}{\textit{Poisson model with fixed effects}} \\
0.1 & 6558 & 809 (672, 964) & 865 (708, 1024) &  55 (-164, 285) & 0.415 (0.409, 0.422) & 0.420 (0.414, 0.427) & 0.0047 (-0.0043, 0.0138) \\ 
0.2 & 6534 & 808 (668, 949) & 875 (722, 1043) &  67 (-170, 277) & 0.416 (0.409, 0.422) & 0.420 (0.414, 0.426) & 0.0045 (-0.0045, 0.0138) \\ 
0.3 & 6516 & 806 (672, 951) & 870 (718, 1040) &  63 (-155, 287) & 0.416 (0.409, 0.422) & 0.420 (0.414, 0.426) & 0.0044 (-0.0049, 0.0135) \\ 
0.4 & 6504 & 805 (675, 947) & 872 (714, 1038) &  66 (-156, 283) & 0.416 (0.409, 0.422) & 0.420 (0.413, 0.426) & 0.0043 (-0.0046, 0.0139) \\ 
0.5 & 6493 & 806 (678, 956) & 880 (722, 1050) &  74 (-165, 285) & 0.416 (0.409, 0.422) & 0.420 (0.413, 0.426) & 0.0041 (-0.0050, 0.0134) \\ 
0.6 & 6489 & 806 (671, 957) & 891 (719, 1065) &  85 (-153, 313) & 0.416 (0.409, 0.422) & 0.420 (0.414, 0.426) & 0.0042 (-0.0043, 0.0135) \\ 
0.7 & 6474 & 803 (666, 946) & 912 (722, 1080) & 109 (-159, 320) & 0.416 (0.409, 0.422) & 0.420 (0.413, 0.426) & 0.0042 (-0.0052, 0.0129) \\ 
0.8 & 6474 & 803 (669, 950) & 926 (720, 1090) & 123 (-162, 342) & 0.416 (0.409, 0.422) & 0.420 (0.414, 0.426) & 0.0042 (-0.0050, 0.0133) \\ 
0.9 & 6480 & 805 (673, 955) & 1166 (716, 1109) & 360 (-156, 373) & 0.416 (0.409, 0.422) & 0.420 (0.414, 0.426) & 0.0042 (-0.0049, 0.0129) \\ 
\multicolumn{8}{l}{\textit{Poisson model with random effects}} \\
0.1 & 5165 & 789 (662, 945) & 861 (705, 1029) &  72 (-149, 300) & 0.416 (0.409, 0.422) & 0.421 (0.415, 0.427) & 0.0047 (-0.0045, 0.0134) \\ 
0.2 & 5179 & 790 (655, 932) & 863 (700, 1023) &  73 (-153, 291) & 0.416 (0.409, 0.423) & 0.421 (0.414, 0.427) & 0.0046 (-0.0039, 0.0140) \\ 
0.3 & 5152 & 794 (657, 942) & 874 (709, 1044) &  79 (-153, 306) & 0.416 (0.409, 0.422) & 0.421 (0.414, 0.427) & 0.0046 (-0.0042, 0.0139) \\ 
0.4 & 5140 & 793 (651, 942) & 884 (706, 1047) &  91 (-151, 318) & 0.416 (0.409, 0.422) & 0.421 (0.415, 0.427) & 0.0047 (-0.0044, 0.0133) \\ 
0.5 & 5125 & 795 (659, 951) & 882 (710, 1053) &  87 (-166, 317) & 0.416 (0.409, 0.423) & 0.421 (0.414, 0.427) & 0.0046 (-0.0046, 0.0134) \\ 
0.6 & 5122 & 799 (655, 949) & 925 (706, 1077) & 126 (-158, 360) & 0.416 (0.409, 0.423) & 0.421 (0.414, 0.426) & 0.0045 (-0.0047, 0.0132) \\ 
0.7 & 5137 & 812 (653, 962) &1065 (712, 1101) & 252 (-181, 369) & 0.416 (0.409, 0.422) & 0.420 (0.414, 0.427) & 0.0045 (-0.0044, 0.0139) \\ 
0.8 & 5133 & 846 (650, 957) &1227 (713, 1137) & 381 (-203, 424) & 0.416 (0.409, 0.422) & 0.421 (0.414, 0.427) & 0.0046 (-0.0047, 0.0134) \\ 
0.9 & 5135 & 1136 (652, 964) &1747 (701, 1177) &611 (-227, 486) & 0.416 (0.409, 0.422) & 0.420 (0.415, 0.426) & 0.0046 (-0.0045, 0.0133) \\ 
\hline
\end{tabular} 
\end{adjustbox} \\
DIC*: This reports the DIC values for the first cost component of the models. Costs are measured using British pound (£). 95\% CIs are reported in parentheses. CI = Credible interval; Con = Control; Int = Intervention; QALY = Quality-adjusted life-year.
\end{sidewaystable}
\newpage

\section{Prior Sensitivity Analysis Under MAR} \label{appendix_prior}
Table~\ref{appendix:table6} presents the sensitivity of the Bayesian cost-effectiveness results, expressed as mean costs and QALYs for each treatment arm along with their 95\% credible intervals, to the prior distributions. Slight modifications were made to the values of the hyperparameters in the prior distributions. The corresponding cost-effectiveness results are similar to those in the main analysis.

\begin{sidewaystable}
\caption{Prior sensitivity analysis of item-level models} \label{appendix:table6}
\footnotesize
\begin{tabular}{p{3cm}p{3cm}p{2cm}p{2cm}p{2cm}p{3cm}p{3cm}p{3cm}}
\hline
Model & Prior & Costs(Con) & Costs(Int) & Incremental Costs & QALYs(Con) & QALYs(Int) & Incremental QALYs \\ \hline
\multicolumn{8}{l}{\textit{Prior on coefficients}} \\
\multirow[t]{4}{3cm}{Normal} & \multirow[t]{2}{3cm}{$\text{Normal}(0,100^2)$} & 705 & 795 & 90 & 0.416 & 0.422 & 0.0058  \\ 
& & (602, 807) & (673, 920) & (-70, 249) & (0.410, 0.422) & (0.416, 0.427) & (-0.0021, 0.0135) \\ 
& \multirow[t]{2}{3cm}{$\text{Normal}(0,1000^2)$} & 706 & 798 & 92 & 0.416 & 0.422 & 0.0058  \\ 
& & (604, 808) & (671, 919) & (-65, 257) & (0.410, 0.422) & (0.416, 0.427) & (-0.0020, 0.0137) \\ 
\multirow[t]{4}{3cm}{Beta Gamma} & \multirow[t]{2}{3cm}{$\text{Normal}(0,100^2)$} & 826 & 892 & 65 & 0.417 & 0.422 & 0.0051  \\ 
& & (680, 981) & (731, 1060) & (-159, 299) & (0.410, 0.423) & (0.416, 0.427) & (-0.0036, 0.0141) \\ 
& \multirow[t]{2}{3cm}{$\text{Normal}(0,1000^2)$} & 824 & 889 & 65 & 0.417 & 0.422 & 0.0051  \\ 
& & (681, 985) & (731, 1059) & (-156, 303) & (0.410, 0.423) & (0.416, 0.427) & (-0.0038, 0.0138) \\ 
\multirow[t]{4}{3cm}{Poisson with fixed effects} & \multirow[t]{2}{3cm}{$\text{Normal}(0,100^2)$} & 806 & 880 & 74 & 0.416 & 0.420 & 0.0041  \\ 
& & (678, 956) & (722, 1050) & (-165, 285) & (0.409, 0.422) & (0.413, 0.426) & (-0.0046, 0.0134) \\ 
& \multirow[t]{2}{3cm}{$\text{Normal}(0,10^2)$} & 804 & 874 & 70 & 0.416 & 0.420 & 0.0044  \\ 
& & (669, 946) & (723, 1055) & (-157, 304) & (0.409, 0.422) & (0.414, 0.426) & (-0.0050,  0.0134) \\ 
\multirow[t]{4}{3cm}{Poisson with random effects} & \multirow[t]{2}{3cm}{$\text{Normal}(0,100^2)$} & 795 & 882& 87 & 0.416 & 0.421 & 0.0046  \\ 
& & (659, 951) & (710, 1053) & (-166, 317) & (0.409, 0.422) & (0.413, 0.426) & (-0.0046, 0.0134) \\ 
& \multirow[t]{2}{3cm}{$\text{Normal}(0,10^2)$} & 795 & 883 & 88 & 0.416 & 0.421 & 0.0048  \\ 
& & (653, 943) & (707, 1045) & (-160, 306) & (0.409, 0.422) & (0.414, 0.427) & (-0.0036, 0.0142) \\ 
\multicolumn{8}{l}{\textit{Prior on standard deviation}} \\
\multirow[t]{4}{3cm}{Normal} & \multirow[t]{2}{3cm}{$\text{Uniform}(0,1000)$} & 705 & 795 & 90 & 0.416 & 0.422 & 0.0058  \\ 
& & (602, 807) & (673, 920) & (-70, 249) & (0.410, 0.422) & (0.416, 0.427) & (-0.0021, 0.0135) \\ 
& \multirow[t]{2}{3cm}{$\text{Uniform}(0,100)$} & 705 & 797 & 92 & 0.416 & 0.422 & 0.0058  \\ 
& & (601, 807) & (668, 921) & (-82, 246) & (0.410, 0.422) & (0.416, 0.427) & (-0.0021, 0.0135) \\ 
\multirow[t]{4}{3cm}{Beta Gamma} & \multirow[t]{2}{3cm}{$\text{Uniform}(0,1000)$} & 826 & 892 & 65 & 0.417 & 0.422 & 0.0051  \\ 
& & (680, 981) & (731, 1060) & (-159, 299) & (0.410, 0.423) & (0.416, 0.427) & (-0.0036, 0.0141) \\ 
& \multirow[t]{2}{3cm}{$\text{Uniform}(0,100)$} & 829 & 884 & 56 & 0.417 & 0.422 & 0.0051  \\ 
& & (684, 983) & (725, 1055) & (-173, 282) & (0.410, 0.423) & (0.416, 0.427) & (-0.0035, 0.0140) \\ 
\hline
\end{tabular} \\
 Costs are measured using British pound (£). 95\% CIs are reported in parentheses. CI = Credible interval; Con = Control; Int = Intervention; QALY = Quality-adjusted life-year.
\end{sidewaystable}

\newpage
\section{Posterior Predictive Checks} \label{appendix_ppc}
\begin{figure}[htp]
       {\centering
       \includegraphics[width=0.8\textwidth]{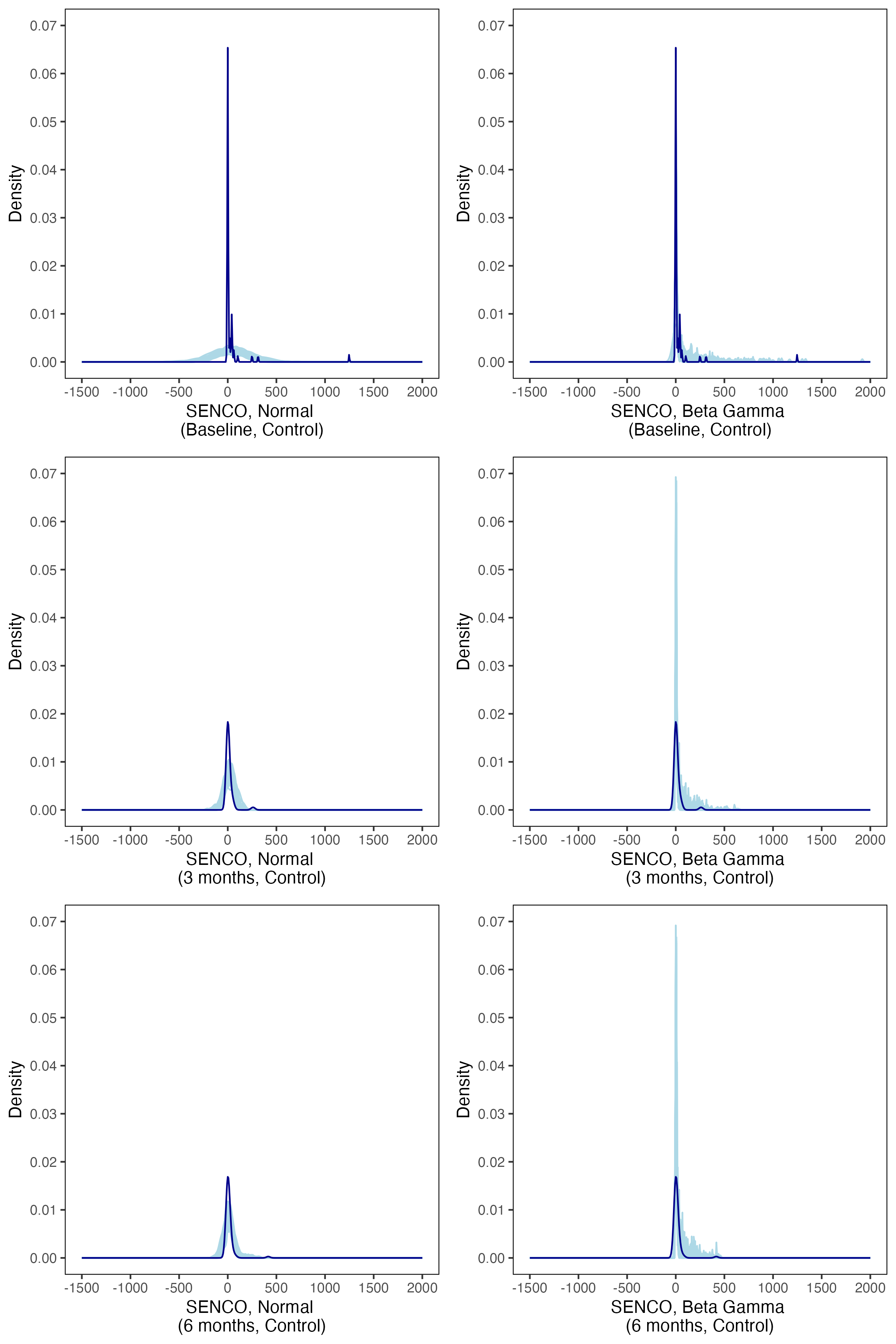}
       \caption{Posterior predictive checks of Normal and Beta-Gamma models for SENCO resource use in the control arm at baseline.}
       \label{appendix:figure1}
       }
       {\small Distribution of replicated SENCO costs data drawn from the posterior predictive distribution and the distribution of observed data estimated using kernel density estimation in control arm for Normal and Beta-Gamma model. In each sub-figure, the dark blue curve represents the empirical distribution of observed data and the light blue curves indicate the distributions of 100 replicated dataset drawn from the posterior predictive distribution. SENCO = Special education need co-ordinator. }
\end{figure}

\begin{figure}[tp]
       {\centering
       \includegraphics[width=1\textwidth]{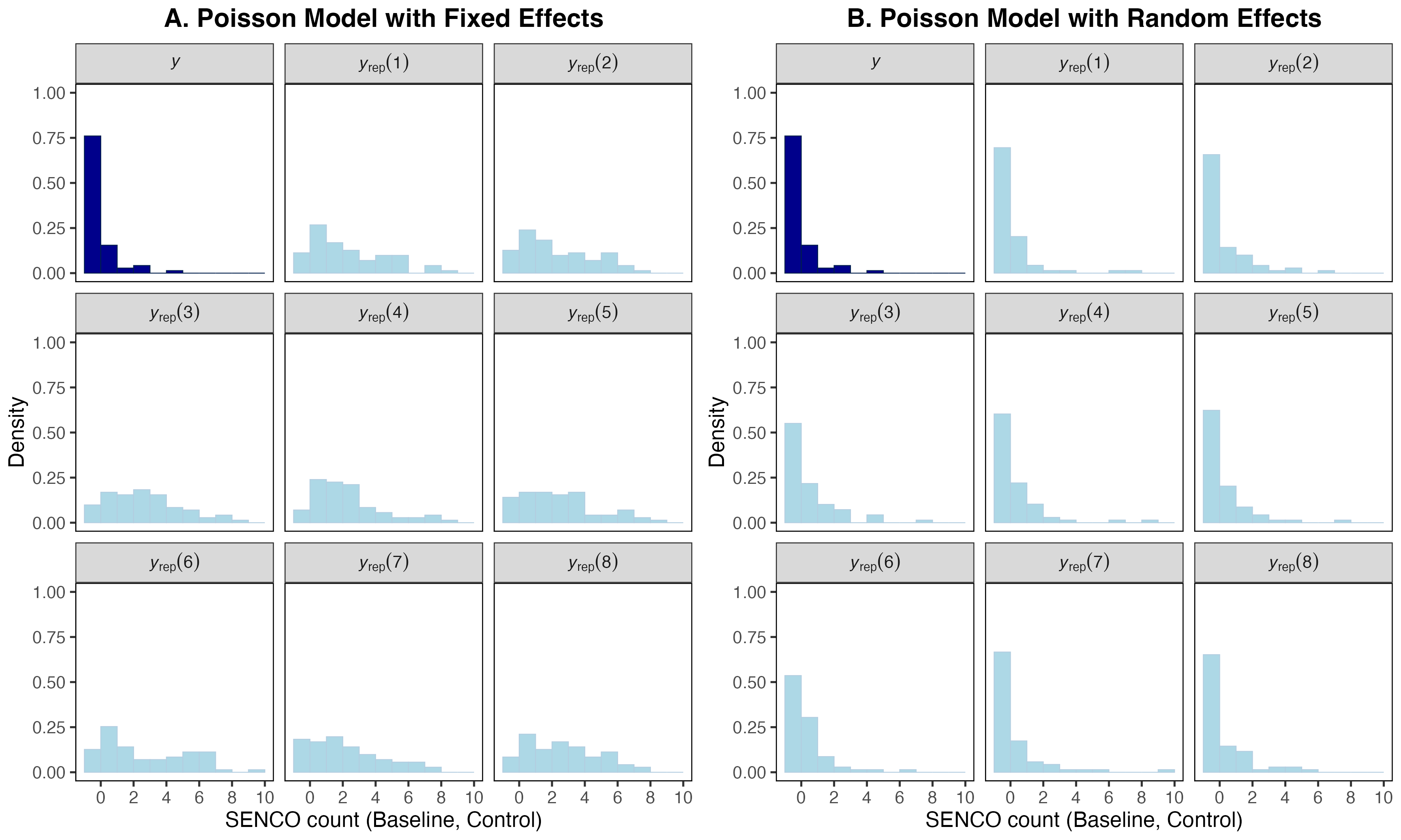}
       \caption{Posterior predictive checks of Poisson models for SENCO resource use in the control arm at baseline.}
       \label{appendix:figure2}
       }
       {\small The observed distribution is compared with eight replicated datasets drawn from the posterior predictive distribution of Poisson models. Panel A shows results for the Poisson model with fixed effects, and Panel B for the Poisson model with random effects. Dark bars represent the observed data and light bars represent replicated data. Similarity between observed and replicated distributions indicates adequate model fit. SENCO = Special education need co-ordinator.}
\end{figure}

We present two figures for posterior predictive checks using SENCO costs as an illustration. 
For item-level cost models, we compare the distributions of replicated data drawn from the posterior predictive distribution with the empirical distribution of the observed data (Figure~\ref{appendix:figure1}). For item-level resource use models, we compare histograms of eight replicated datasets with the histogram of the observed data (Figure~\ref{appendix:figure2}).

\newpage
\section{Sensitivity Analysis: Missing Not at Random (MNAR)} \label{appendix_mnar}
\begin{figure}[htp]
    \centering
    \includegraphics[width=1\textwidth]{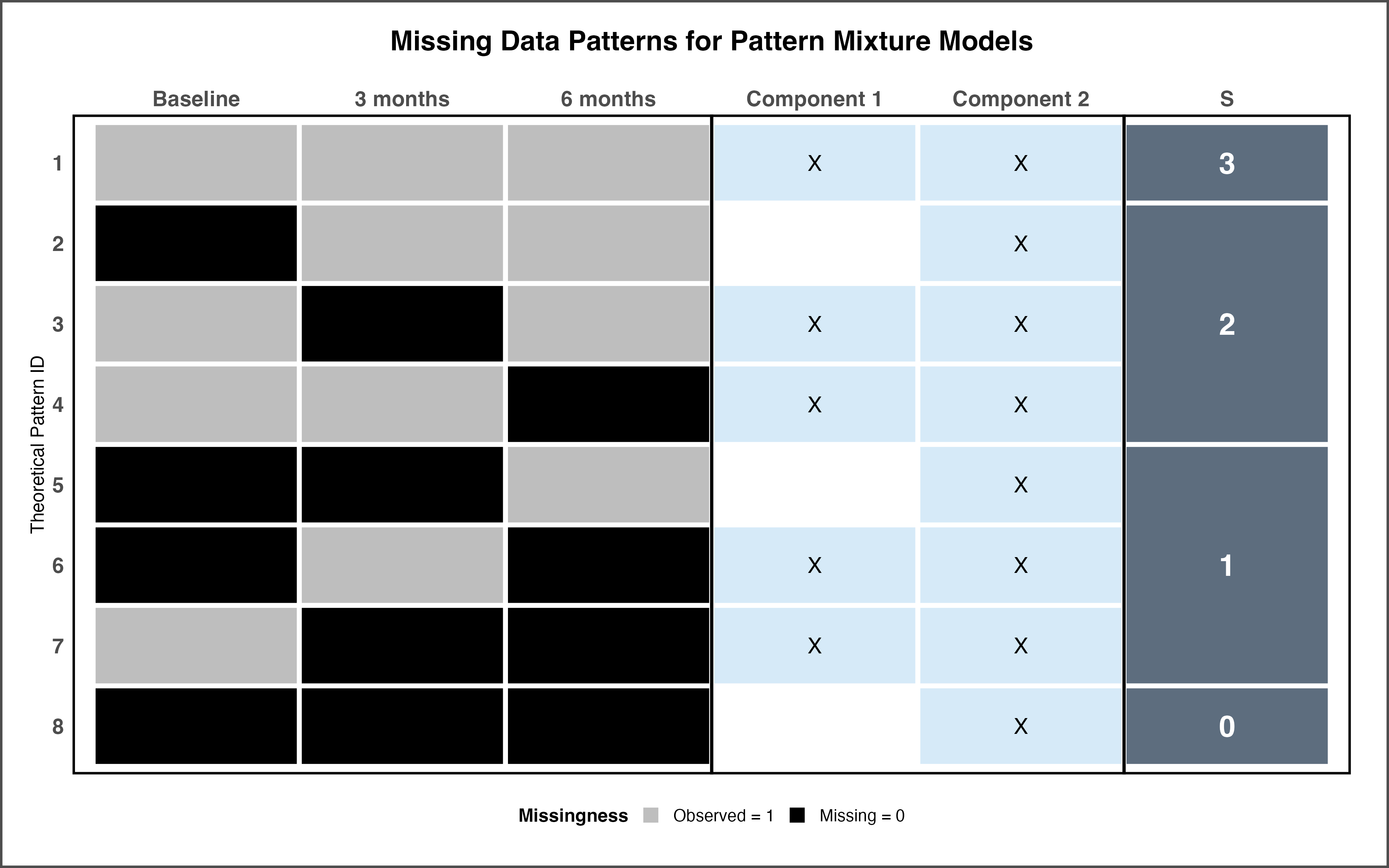}
    \caption{Missing data patterns for pattern mixture models}
    \label{appendix:figure3}
    {\small \raggedright
    $S = \sum_{j=1}^3 r_{ijk}$ where $S \in \{0,1,2,3\}$. X indicates the missing pattern is covered by the model component.
    \par}
\end{figure}

Figure~\ref{appendix:figure3} depicts the theoretical missing data patterns and their realisation in the ORBIT trial. It can be viewed as a simplified version of Figure \ref{fig1} and is used to support the development of a pattern-mixture model for MNAR missing cost items.

\begin{figure}[htp]
    \centering
    \includegraphics[width=1\textwidth]{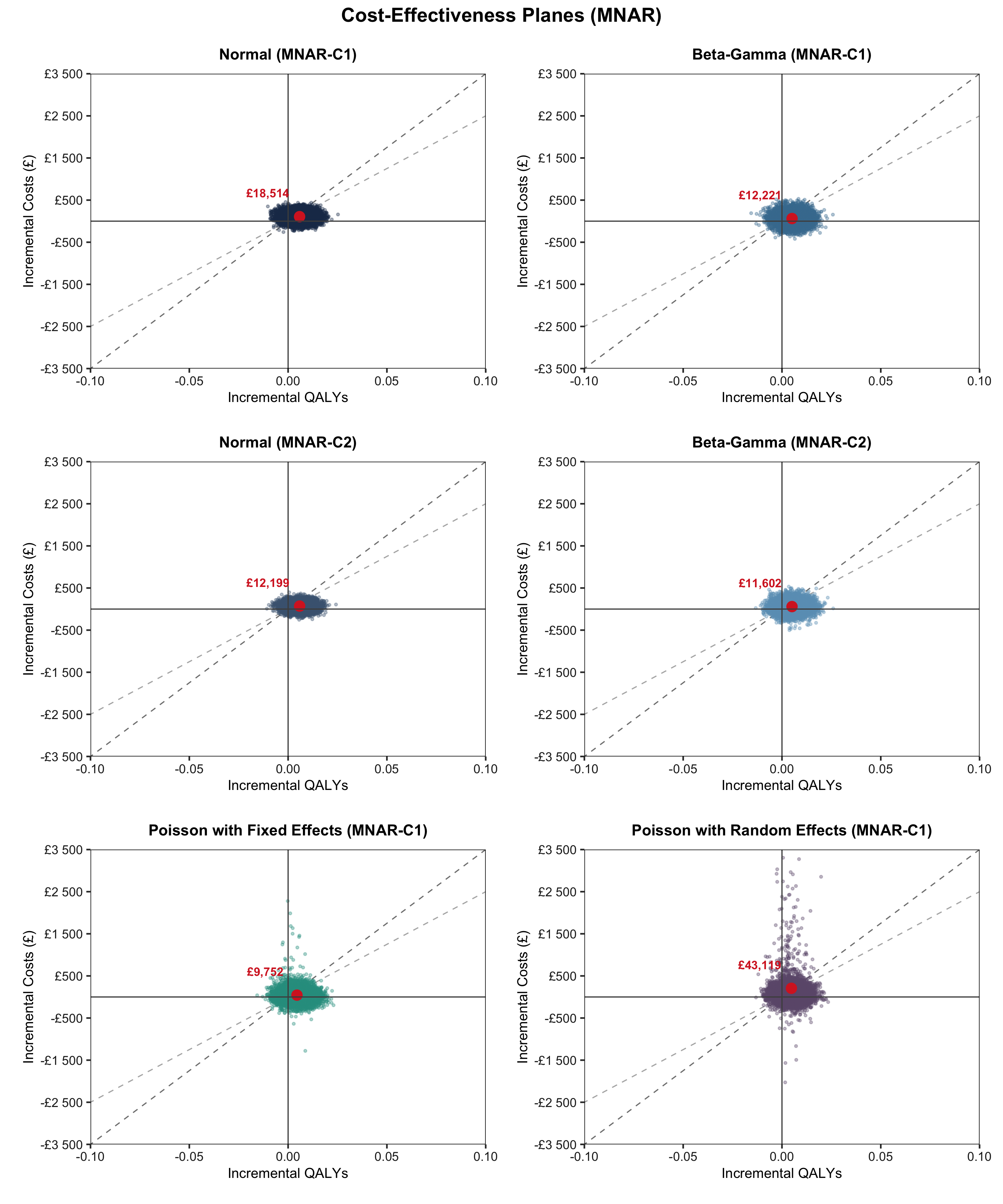}
    \caption{Cost-Effectiveness Planes (MNAR)}
    \label{appendix:figure4}
    {\small \raggedright
    Cost-effectiveness acceptability curves of item-level models under MNAR. MNAR = Missing not at random. The solid lines represent the Normal and Beta–Gamma models, and the dashed lines represent the Poisson models with fixed and random effects. The two dashed vertical lines indicate willingness-to-pay thresholds of £25,000 and £35,000 per QALY gained.
    \par}
\end{figure}

Figure~\ref{appendix:figure4} presents the cost-effectiveness planes of Normal, Beta-Gamma and Poisson models under MNAR.

\end{appendices} 
\end{document}